\documentclass{svjour3}
\usepackage{graphicx}
\usepackage{float}
\usepackage[justification=centering]{subfig}
\usepackage{cite}
\usepackage{hyperref}
\hypersetup{hidelinks}
\usepackage{url}
\usepackage{listings}
\usepackage{xcolor}
\usepackage{tcolorbox}
\usepackage{caption}
\usepackage{booktabs}
\usepackage{multirow}
\usepackage{enumerate}
\usepackage{textcomp}
\usepackage{booktabs}
\usepackage[sort,compress]{natbib}
\usepackage{diagbox}
\usepackage{enumitem}
\usepackage[misc,geometry]{ifsym}
\usepackage{fontawesome5}
\usepackage{footnote}

\newtcolorbox{qoutebox}[3][]
{
  colframe = gray!30!white,
  colback  = #2!10,
  #1,
}

\begin{document}

\title{Demystifying Code Snippets in Code Reviews: A Study of the OpenStack and Qt Communities and A Practitioner Survey}
\titlerunning{Code Snippets in Code Reviews}

\author{Beiqi Zhang   \and
        Liming Fu     \and
        Peng Liang    \and
        Jiaxin Yu     \and
        Chong Wang
}

\institute{Beiqi Zhang \and Liming Fu \and Peng Liang (\Letter) \and Jiaxin Yu \and Chong Wang (\Letter) \at School of Computer Science, Wuhan University, Wuhan, China \\ 
            Hubei Luojia Laboratory, Wuhan, China\\
            \email{\{zhangbeiqi, limingfu, liangp, jiaxinyu, cwang\}@whu.edu.cn}
}

\date{Received: date / Accepted: date}

\maketitle

\begin{abstract}
Code review is widely known as one of the best practices for software quality assurance in software development. In a typical code review process, reviewers check the code committed by developers to ensure the quality of the code, during which reviewers and developers would communicate with each other in review comments to exchange necessary information. As a result, understanding the information in review comments is a prerequisite for reviewers and developers to conduct an effective code review. Code snippet, as a special form of code, can be used to convey necessary information in code reviews. For example, reviewers can use code snippets to make suggestions or elaborate their ideas to meet developers' information needs in code reviews. However, little research has focused on the practices of providing code snippets in code reviews.
To bridge this gap, we conduct a mixed-methods study to mine information and knowledge related to code snippets in code reviews, which can help practitioners and researchers get a better understanding about using code snippets in code review. Specifically, our study includes two phases: mining code review data and conducting practitioners' survey. In Phase 1, we conducted an exploratory study to mine code review data from two popular developer communities (i.e., OpenStack and Qt). We manually labelled 69,604 review comments and finally identified 3,213 review comments that contain code snippets. Based on the code review data collected, we analyzed the extent of using code snippets, the reviewers' purposes of providing code snippets, the developers' acceptance of code snippet suggestions, and the reasons that developers do not accept code snippet suggestions in code reviews. In Phase 2, we used an online questionnaire to survey practitioners from industry. By analyzing the 63 valid responses we received, we explored the scenarios reviewers provide code snippets, the developers' attitudes towards code snippets, and the characteristics of code snippets developers expect reviewers to provide in code reviews.
Our results show that: (1) code snippets are not frequently used in code reviews, and most of the code snippets are provided by reviewers rather than developers; (2) the purposes of reviewers providing code snippets in code reviews are \textit{Suggestion} and \textit{Citation}, in which \textit{Suggestion} is the main purpose; (3) most developers would accept reviewers' code snippet suggestions; (4) the most common reasons that developers do not accept reviewers' code snippet suggestions in code reviews are \textit{difference in the opinions between developers and reviewers} and \textit{reviewer's suggestion is flawed}; (5) reviewers often provide code snippets in code reviews \textit{when code is more illustrate than words}; (6) most developers hold positive attitudes towards code snippet comments; and (7) most developers expect that code snippets in review comments are \textit{understandable} and \textit{fitting into existing code}. The study results highlight that reviewers can provide code snippets in appropriate scenarios to meet developers' specific information needs in code reviews, which will facilitate and accelerate the code review process.
\end{abstract}

\section{Introduction}
\label{sec:introduction}

Code review has proven to be one of the best practices in software development which is conducted by the developers other than the code author through a manual inspection of source code \citep{ackerman1989software}. Modern Code Review (MCR) is a type of code review that is informal, tool-based, and occurs regularly in practice \citep{bacchelli2013expectations}. Nowadays, MCR practices are becoming more lightweight, co-located, continuous, and asynchronous \citep{sadowski2018modern, badampudi2023modern}, due to the rise of agile and Open Source Software (OSS) development \citep{rigby2013convergent}. In a typical MCR process, reviewers and developers engage in asynchronous online discussions to exchange information with each other, ensuring that the proposed code changes are of sufficient quality and align with the direction of the project before they are accepted \citep{pascarella2018information}. MCR not only provides assurance of software quality \citep{davila2021systematic}, but also contributes to design improvement, knowledge sharing, and code ownership \citep{nazir2020modern}. 

Understanding the information in code review comments is a prerequisite of efficient code review process. When developers submit code changes, they are concerned about information related to whether they make mistakes in their new code and whether they follow their team's coding conventions \citep{ko2007information}, which can be obtained from code review comments. However, it is challenging for developers to get the necessary information in MCR \citep{bacchelli2013expectations}. Although many tools (e.g., Gerrit, Veracode, Reshift) support the process of MCR, developers and reviewers still have the need of richer communications including a wide range of mechanisms in code review process \citep{bacchelli2013expectations}. \cite{sutherland2009can} investigated the code review practices of software product teams at Microsoft to understand the nature of the code review dialog and exchange of information, how the information was retained, and the nature of its later reuse. The results indicated that the retention and recovery of information in code reviews is not well supported in the current environment.

Code review comments are text-based, which may contain textual content like URL links and code snippets for developers to get useful information for a better understanding of the review comments. Previous studies have investigated the practice of link sharing and their purposes in code reviews, and have explored what types of information could be provided through link sharing \citep{wang2021understanding}. Similar to links, code snippets could also convey necessary information for developers in code reviews. However, the different nature of links and code snippets leads to their dedicated purposes and influences in the practice of code review. To the best of our knowledge, little study has investigated the use of code snippets in MCR, and it is still unknown about the purposes of providing code snippets and the practices and knowledge of using code snippets in code reviews.

To bridge this gap, we conducted this mixed-methods study to provide a comprehensive understanding of code snippets in code reviews. We mined code review discussions from four most active projects selected from the OpenStack\footnote{\url{https://www.openstack.org/}} community (Nova\footnote{\url{https://wiki.openstack.org/wiki/Nova}} and Neutron\footnote{\url{https://wiki.openstack.org/wiki/Neutron}}) and the Qt\footnote{\url{https://www.qt.io/}} community (Qt Base\footnote{\url{https://github.com/qt/qtbase}} and Qt Creator\footnote{\url{https://www.qt.io/product/development-tools}}) based on the number of closed code changes. The code review process of these four projects are all supported by Gerrit\footnote{\url{https://www.gerritcodereview.com}}, a Web-based code review platform built on the top of Git. 
We also used an online survey to explore code snippets in code reviews by following the guidelines provided by Kitchenham and Pfleeger \citep{kitchenham2008personal}. We sent the survey questionnaire to potential respondents from the popular developer groups (i.e., LinkedIn) and to practitioners from the four selected projects in OpenStack and Qt (i.e., Nova, Neutron, Qt Base, and Qt Creator). In total, we got 3,197 review comments containing code snippets by manually checking 69,604 code review discussions obtained from the mined data sources, and we received 63 valid responses from the survey questionnaire. A comprehensive quantitative and qualitative analysis were conducted to study the extent of using code snippets, the purposes of providing code snippets, and how developers treat code snippet suggestions in code reviews, in order to understand the practices and purposes of code snippets in MCR. Our results suggest that: (1) Code snippets are not prevalently used in code reviews, and most of the code snippets in review comments are provided by reviewers. (2) Reviewers make code snippet suggestions in code review with the aim of \textit{Suggestion} and \textit{Citation}, in which \textit{Suggestion} is the main purpose. (3) For code snippet suggestions, most developers would accept them. (4) The main reasons why developers not accept reviewers' code review suggestions are \textit{Difference in the opinions between developers and reviewers} and \textit{Reviewer's suggestion is flawed}. (5) Reviewers tend to provide code snippets in code reviews \textit{when code is more illustrate than words}. (6) For review comments containing code snippets, most developers hold positive attitudes. (7) Developers expect that code snippets provided by reviewers in code reviews can be \textit{Concise}, \textit{Correct}, and \textit{Executable}.

This paper extends our earlier work on studying code snippets in code reviews \citep{fu2022understanding} through the following additions:
\begin{enumerate}[itemindent=1.5em]
    \item We extended our dataset by including the code review data from two most active projects of the Qt community and collecting all the closed code changes updated in 2021, which are 127,182 code review comments in total.
    \item We explored the reasons why developers do not accept code snippet suggestions in code reviews (RQ4).
    \item We further investigated the scenarios when reviewers provide code snippets, developers' attitude towards code snippets, and the characteristics of code snippets that developers expect reviewers to provide in code reviews (RQ5, RQ6, and RQ7) through an industrial survey.
\end{enumerate}

The rest of this paper is structured as follows: Section \ref{sec:relatedWork} presents the related work. Section \ref{sec:methodology} describes the research design of this study. Section \ref{sec:results} provides the study results, which are further discussed in Section \ref{sec:discussion}. The potential threats to validity are clarified in Section \ref{sec:threats}, and Section \ref{sec:conclusions} concludes this work with future directions.

\section{Related Work}
\label{sec:relatedWork}

\subsection{Code Snippet}
Several studies have focused on identifying various information from code snippets. \cite{subramanian2013making} performed static analysis on code snippets from accepted answers in Stack Overflow to understand that the structural information could be obtained from code snippets, and they used such information to effectively identify API usage in snippets. Their code snippet analysis approach could dramatically improve the accuracy of identifying structural relationships from code snippets compared to lexical approaches. \cite{chatterjee2017what} conducted an exploratory study to extract what kinds of information associated with code snippets are available in various software-related documents, including blog posts, API documentation, code reviews, and public chats. Their results revealed the characteristics of the code snippets embedded in different document types, the kinds of information contained in code snippets, and what cues can indicate code snippet related information. \cite{panichella2012mining} mined code snippet descriptions from developer communities such as mailing lists and bug reports. They used such information to propose an approach which could automatically extract method descriptions in developers' communications in order to help developers understand the code. The method descriptions extracted by their approach reached a high precision. 

Many studies have focused on tools related to code snippets. \cite{galenson2014codehint} presented an efficient tool called CodeHint to help synthesize code snippets that use real-world Java features, and their results show that the algorithms of CodeHint could significantly improve developers' productivity. \cite{wong2013autocomment} mined question and answer platforms like Stack Overflow which contain code descriptions written by developers to propose AutoComment tool, which could automatically generate code snippet comments. They applied Natural Language Processing (NLP) techniques in this code-description generation tool to analyze sentence semantics, and many practitioners thought that the generated comments by the tool were accurate, adequate, concise, and useful in helping them understand the related code. \cite{campbell2017nlp2code} proposed a tool called NLP2Code integrated in Eclipse IDE which has a content assist feature for code snippets. NLP2Code could recommend relevant code snippets on Stack Overflow based on the natural language typed by the developers, thus saving developers' time of searching the Web for the required code snippets.


Compared to the existing work related to code snippets discussed above, our work tends to focus on the practice of code snippets in code reviews by investigating the extent of using code snippets, the purposes of providing code snippets, and how developers treat code snippet suggestions in code reviews.

\subsection{Code Review}
Code review is a mature practice and essential part in modern software development. In recent years, many studies have explored the modern code review process in practice. Some studies chose to investigate code review process based on pull requests. \cite{zampetti2019study} investigated how developers use the outcome of Continuous Integration (CI) builds during modern code review based on the discussion of pull requests. Their results show that while pull requests with passed builds have slightly more chances of being merged than when builds are broken, other process-related factors have a stronger correlation with such a merger. \cite{wessel2020what} explored why open source maintainers integrate code review bots into the pull request workflow and how they perceive the changes these bots induce. They found that the most frequently mentioned motivations for using bots include automating repetitive tasks, improving tools' feedback, and reducing maintenance effort. Some studies chose to use code review tools to investigate code reviews. \cite{mcintosh2016empirical} mined Gerrit review database to study the relationships between post-release defects and modern code review. They found that code review coverage, participation, and expertise have a clear link with software quality, and their results empirically confirm the intuition that code which has not been well reviewed has a negative impact on software quality in large system development.

\cite{pascarella2018information} gathered reliable data from three large open-source software projects on reviewers' information needs. Based on their results, they found that the most important needs in code reviews are the needs to know whether a proposed alternative solution is valid and whether the understanding of the reviewer about the code under review is correct. \cite{ko2007information} analyzed software developers’ day-to-day information needs in collocated software development teams. Their results show that the most often deferred searches included knowledge about design and program behavior (e.g., why code was written a particular way and what a program was supposed to do). \cite{sutherland2009can} performed an investigation of the code review practices of software product teams at Microsoft to better understand the nature of the code review dialog and exchange of information, how the information was retained, and the nature of its later reuse. Their results reveal that code reviews in collocated development environments such as Microsoft use a mix of face-to-face and electronic communication. 

\cite{hirao2022code} studied patches with divergent review scores in the OpenStack and Qt communities. Their results suggest that review tooling should integrate with release schedules and detect concurrent development of similar patches to optimize review discussions with divergent scores. \cite{cunha2007does} investigated if there is a specific personality type that is correlated with performance on a code-review task. They suggested that software companies should capitalize on the strengths of their employees who can better perform code review tasks than others and consider employees perhaps previously overlooked for particular code review tasks.

Several studies have focused on investigating a variety of artifacts in code review process. \cite{zanaty2018empirical} studied the frequency and nature of design discussions in code reviews to better understand to what extent design is discussed during code review. Their manual analysis indicates that though design-related discussions are still rare during code review process, design related comments are constructive and can provide suggestions to mitigate design issues. \cite{fu2022potential} conducted an exploratory study in an attempt to understand the nature of Potential Technical Debt (PTD) in code reviews. Their findings indicate that review-based detection of PTD is one of the trustworthy mechanisms in development. \cite{kashiwa2022empirical} conducted a study aiming to understand the effect of Self-Admitted Technical Debt (SATD) comments on accepting and revising patch-sets and the practice of introducing SATD in code reviews. Their results show that 28\textasciitilde48\% of SATD comments are introduced during MCRs. \cite{han2021understanding} conducted an empirical study to investigate the concept behind code smells identified in code reviews and what actions reviewers suggest and developers take in response to the identified smells, and they found that the majority of smell-related suggestions were accepted by developers.

Different from the aforementioned works, our work intends to study the distribution, purposes, and acceptance of code snippets in code reviews by manually extracting code snippet related comments from the most active projects of the OpenStack and Qt communities, and the scenarios in which reviewers provide code snippets, the developers' attitudes towards code snippets, and the characteristics of code snippets developers expect reviewers to provide in code reviews from an industrial survey.


\section{Methodology}
\label{sec:methodology}

\begin{figure}[htbp]
	\centering
	\includegraphics[width=1.0\linewidth]{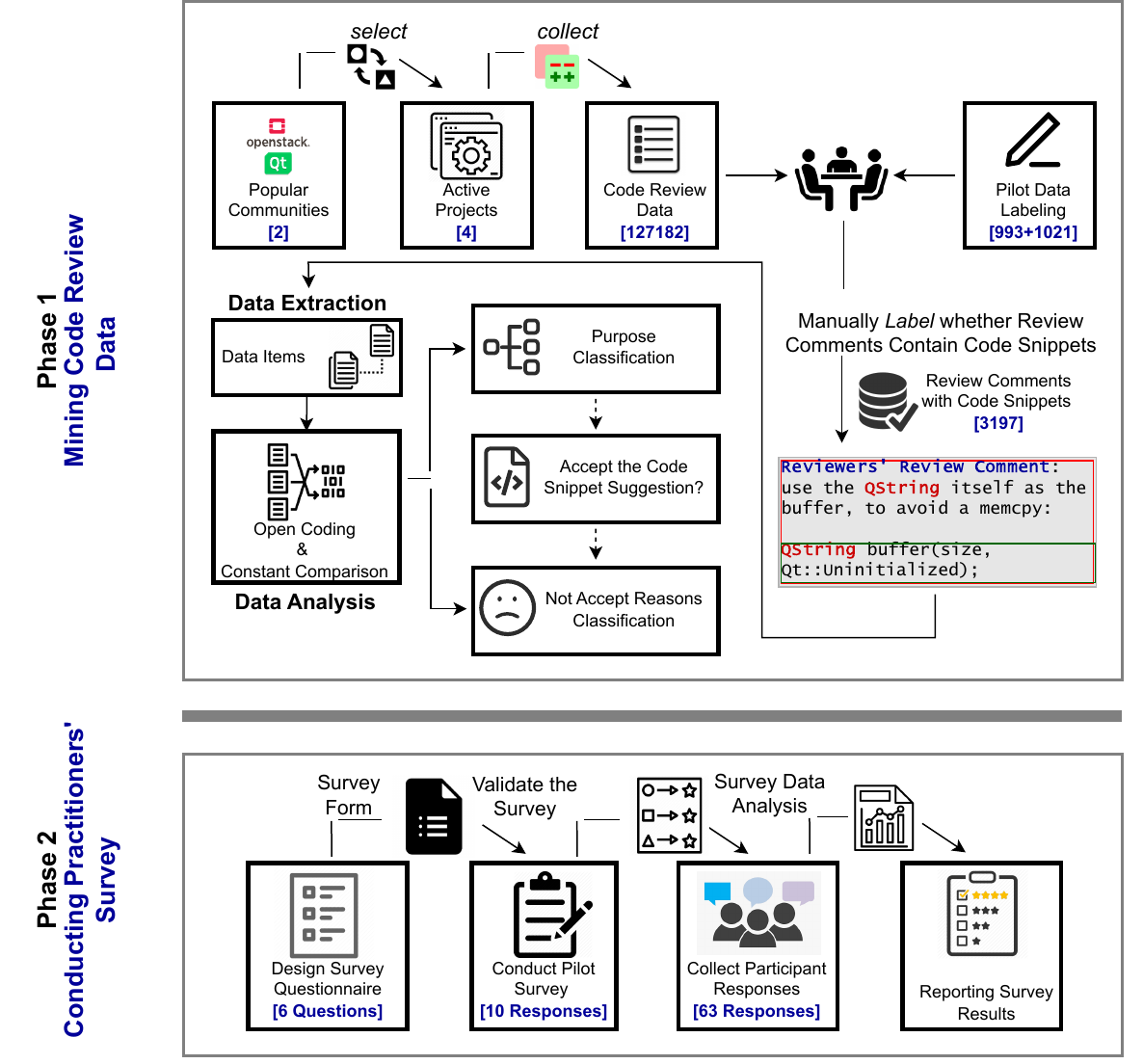}
	\caption{Overview of the mixed-methods research process}
	\label{fig: Overview of the mixed-methods research process}
\end{figure}

We used a mixed-methods approach that combines an exploratory study through mining code review data from the OpenStack and Qt communities and an industrial survey with 63 practitioners. In this section, we first present our seven Research Questions (RQs). Then we describe the research process (see Fig.~\ref{fig: Overview of the mixed-methods research process}) and detail the methods used to collect, label, extract, and analyze the data in this study. 


\subsection{Research Questions}
\label{subsec:research_questions}
The goal of this study is to understand the practices of using code snippets in the context of code review, and to develop a practical guideline of using code snippets in code reviews. This goal can be further decomposed into seven RQs as listed below. Specifically, the first four RQs (i.e., RQ1, RQ2, RQ3, and RQ4) are answered through an exploratory study by analyzing the code review data collected from the OpenStack and Qt communities. The results of these four RQs can help understand the extent of using code snippets in code reviews, the reviewers' purposes of providing code snippets in code reviews, the developers' acceptance of code snippets in code reviews, and the reasons developers do not accept code snippet suggestions in code reviews. 
The last three RQs (RQ5, RQ6, and RQ7) are answered through an industrial survey of 63 practitioners, in order to understand the scenarios in which reviewers provide code snippets, developers' attitudes towards code snippets in code reviews, and the characteristics that developers want code snippets to have in code reviews.

\noindent\textbf{RQ1: To what extent are code snippets used in code reviews?}

\noindent\textbf{Motivation:} Currently, we do not know how frequently code snippets are used in code reviews in the modern code review process. As an exploratory study on code snippets in code reviews, this RQ aims to explore the distribution and proportion of review comments that contain code snippets. The answer of this RQ can help to get a basic overview of code snippets in code reviews. 

\noindent\textbf{RQ2: What are the purposes of code snippets provided by reviewers in code reviews?}

\noindent\textbf{Motivation:} Previous work has studied the purposes of shared links in code reviews to investigate what kinds of information could be provided through link sharing \citep{wang2021understanding}. Code snippets can also provide important information in code reviews similar to shared links, however, we have no idea about the purposes of providing code snippets in code reviews. According to the results of RQ1, we found that most code snippets in review comments are provided by reviewers. This RQ aims to understand the purposes of code snippets provided by reviewers in code reviews. The answer of this RQ can help to get a better understanding of the roles code snippets play in code review.

\noindent\textbf{RQ3: How do developers treat code snippet suggestions in code reviews?}

\noindent\textbf{Motivation:} The reaction of developers to the review comments that contain code snippets is meaningful to explore the impact of code snippets in code reviews. According to the results of RQ2, we found that the most common purpose for reviewers to provide code snippets in code reviews is \textit{Suggestion} for developers. Developers may take different actions towards reviewers' code snippet suggestions. They could decide whether or not to accept reviewers' code snippet suggestions or just ignore them. This RQ aims to explore how developers treat code snippet suggestions in code reviews by investigating how many code snippet suggestions are accepted, ignored, and not accepted, respectively. Such information can help to understand the usefulness of the code snippets in code reviews.

\noindent\textbf{RQ4: What are the reasons that developers do not accept code snippet suggestions in code reviews?}

\noindent\textbf{Motivation:} During the code review process, not all code snippet suggestions would be accepted by developers. Sometimes, developers may not follow reviewers' code review suggestions even though developers agree with the reviewers, for other reasons. This RQ aims to explore the reasons why developers do not accept reviewers' code snippet suggestions. The answer of this RQ can help reviewers be aware of the reasons behind developers' nonacceptance of the code snippet suggestions and thus help reviewers improve their code reviews and avoid the situations that result in code snippet suggestions not being accepted.

\noindent\textbf{RQ5: In which scenarios do reviewers often provide code snippets in code reviews?}

\noindent\textbf{Motivation:} During the code review process, in some scenarios reviewers tend to provide code snippets in review comments, while in some other scenarios, they tend to provide plain text or shared links instead. This RQ aims to investigate the situations when reviewers provide code snippets in code reviews. The answer of this RQ can help acquire the knowledge and practices of using code snippets in code reviews. Such information can also guide reviewers when to provide code snippets in code reviews.

\noindent\textbf{RQ6: What are the attitudes of developers towards code snippets provided in code reviews?}

\noindent\textbf{Motivation:} Intuitively, developers may have a positive attitude towards code snippets provided in code reviews as code is more convincing and direct than review comments in plain text. This RQ aims to understand the actual attitudes of developers towards code snippets in code reviews. Answering this RQ can help to guide whether reviewers are encouraged to provide code snippet in order to make the discussions during the code review process more smoothly.

\noindent\textbf{RQ7: What are the characteristics of code snippets that developers expect reviewers to provide in code reviews?}

\noindent\textbf{Motivation:} Code snippets with specific characteristics (e.g., code quality and readability) in code reviews may decide whether developers will accept reviewers' suggestions or not. This RQ aims to summarize the characteristics that developers hope code snippets in code reviews to have. The results of this RQ can help reviewers be aware of the key points of providing code snippets in code reviews and thus make their code snippet suggestions more acceptable.

\subsection{Exploratory Study Design}
\subsubsection{Data Collection}
\label{datacollection}
For the exploratory study, we used four active projects from two large and popular open-source communities: OpenStack and Qt. OpenStack is a set of software components that provide common services for cloud infrastructure, and is contributed by many well-known software companies, e.g., IBM, VMware, and NEC \citep{thongtanunam2018review}. Qt is a cross-platform tool and UI framework for creating graphical user interfaces and developing multi-platform applications. In addition to Qt company, many organizations and individuals who use Qt as a development platform also contribute in the open development of Qt through the Qt community. Meanwhile, the OpenStack and Qt communities have made a serious investment in code review for many years \citep{hirao2022code}, and the code review data from these two communities have been widely used in many studies related to code review (e.g., \cite{wang2019why, thongtanunam2016revisiting, ruangwan2019impact, hamasaki2013who, ueda2018how, wang2021understanding}).

Based on the reasons above, we argue that it is appropriate and representative to conduct our code review research based on these two communities. The OpenStack and Qt communities are both composed of a set of projects, and we selected the two most active projects in each community as our subject projects (based on the number of closed code changes): Nova (a cloud computing fabric controller) and Neutron (a networking service) in the OpenStack community and Qt Base (a module which offers classes for embedded Linux devices) and Qt Creator (a cross-platform Integrated Development Environment (IDE)) in the Qt community.

Both OpenStack and Qt communities use Gerrit\footnote{\url{https://www.gerritcodereview.com/}} to support their code review process. Gerrit is a Web-based code review tool built on top of Git, a well-known distributed control system for code. Gerrit provides various REST APIs to acquire code review data. By using the RESTful API provided by Gerrit, we were able to collect all the closed code changes of the OpenStack projects (Nova and Neutron) and the Qt projects (Qt Base and Qt Creator) from 2020 to 2021. Then, we extracted all available review comments for these code changes and stored the review data in local files for further analysis. Table \ref{Overview of the collected code review data for each project} shows the details of the code review data collected from each project. In total, we collected 34,370 code changes and 127,182 review comments from the OpenStack and Qt communities, which were used as the basic code review data for the exploratory study.

\begin{table}[htbp]
\centering
\renewcommand\arraystretch{1.5}
\caption{Overview of the collected code review data from each project}
\label{Overview of the collected code review data for each project}
\begin{tabular}{p{3cm}m{3cm}<{\centering}m{5cm}<{\centering}}
\hline
\textbf{Project} & \textbf{\#Code Changes} & \textbf{\#Review Comments (RCs)} \\ \hline
Nova             & 2,960                           & 20,748                             \\ 
Neutron          & 3,475                           & 13,541                             \\ 
Qt Base          & 17,181                          & 66,878                             \\ 
Qt Creator       & 10,754                          & 26,015                             \\ \hline       
\textbf{Total}   & 34,370                          & 12,7182                            \\ \hline
\end{tabular}
\end{table}

\subsubsection{Data Labelling}
\label{datalabelling}
After collecting the code review data from the OpenStack and Qt communities, we got a total of 127,182 review comments to label if they contain code snippets. It is time-consuming to label such a large number of code review comments manually, and we decided to eliminate some irrelevant code review comments by using certain measures. The whole process of data labelling is divided into the following five steps:

In \textbf{step one}, we removed the review comments that were generated by bots (i.e., Zuul in OpenStack and Qt Sanity Bot in Qt) since we aimed at exploring code snippets in code reviews from the perspective of developers and reviewers.

In \textbf{step two}, considering that Nova and Neutron projects are mainly written in Python (more than 97\%), and Qt Base and Qt Creator projects are mainly written in C++ (more than 85\%), we decided to focus on the review comments that contain Python code snippets in Nova and Neutron and the review comments that contain C++ code snippets in Qt Base and Qt Creator. In addition, we only retained the review comments from the source files written in the main programming language of each project, as these source files are directly related to the code snippets written in corresponding programming languages. In particular, for Nova and Neutron projects, we only kept review comments from the source files with \textsc{.py} file extension, and for Qt Base and Qt Creator projects, we only kept review comments in the source files with \textsc{.cpp} or \textsc{.h} file extension. After the two steps, a part of the irrelevant review comments were filtered out. The counts of remaining review comments for each project are presented in Table \ref{Count of review comments for the projects in OpenStack and Qt communities}.

In \textbf{step three}, the first and second authors manually labelled the remaining review comments after a pilot data labelling. Specifically, the pilot data labelling process is composed of the following substeps: (1) With a 95\% confidence level and a 3\% margin of error \citep{israel1992determining}, the first author randomly selected 993 review comments from the code review data of the OpenStack projects in 2020 and 1,021 review comments from the code review data of the Qt projects in 2021. We randomly selected review comment data from different projects in different years in order to increase the diversity of the pilot labelling data. (2) The first and second authors labelled independently whether the review comments should be included or not. (3) Review comments labelled by the two authors were compared to measure the inter-rater reliability and Cohen's Kappa coefficient \citep{Jacob1960Coefficient} was calculated as a way to verify the consistency on the labelled review comments between the two authors. The Cohen's Kappa coefficient for the OpenStack community is 0.86, while it is 0.92 for the Qt community, both higher than 0.8, thus indicating a high degree of consistency between the two authors.

In the data labelling process, the first two authors followed a set of inclusion and exclusion criteria: (I1) If a review comment contains code snippets (source code or pseudocode) with at least one valid statement, we include it. (E1) If a review comment only contains the name of a certain variable or function, we exclude it. (E2) If the code snippets contained in a review comment come from log files (e.g., error stack trace), we exclude it. To better illustrate the inclusion and exclusion criteria, consider the following two exemplary review comments. The first review comment is included as it contains a valid code statement. The second review comment is excluded. Though it contains a valid code statement, the code snippet is from the system error output log. 


\begin{qoutebox}{white}{}
    \textbf{Link:} \url{http://alturl.com/9gnck}\\
    \textbf{Reviewer:} ``shorter suggestion: \\
\texttt{return parent.isValid() ? treeItemAtIndex(parent)->childCount()
                        : m\_rootItem->childCount();}''
\end{qoutebox}


\begin{qoutebox}{white}{}
    \textbf{Link:} \url{http://alturl.com/e2ib3}\\
    \textbf{Reviewer:} ``src/plugins/qmldesigner/components/itemlibrary/itemlibraryassetimporter.cpp:504: error: `m\_importFiles' was not declared in this scope \\
 \texttt{ 504 |  if (model \&\& !m\_importFiles.isEmpty()) \{ } \\
An ifdef is missing somewhere. Please fix.''
\end{qoutebox}

It is easy to tell whether a review comment contains code snippets or not based on the inclusion and exclusion criteria as most of the review comments we collected are trivial cases. During the data labelling process, if the first two authors were unsure whether or not to include a review comment, which is very rare (only 14), the third author was invited to discuss until an agreement was reached. After manually labelling all the candidates, we finally got a total of 3,213 review comments that contain code snippets from the four projects of the OpenStack and Qt communities. Table \ref{Count of review comments for the projects in OpenStack and Qt communities} shows the count of review comments after each step of data labelling. Note that we used the data after \textbf{step two} to answer RQ1, and the data after \textbf{step three} to answer RQ2.

\begin{table}[htbp]
\renewcommand\arraystretch{1.5}
\centering
\caption{Counts of review comments of the four projects in the OpenStack and Qt communities}
\label{Count of review comments for the projects in OpenStack and Qt communities}
\begin{tabular}{p{1.9cm}m{1.9cm}<{\centering}m{3.4cm}<{\centering}m{3.5cm}<{\centering}}
\hline
\textbf{Project} & \textbf{\#RCs}         & \textbf{\#RCs after Step Two}         & \textbf{\#RCs after Step Three}  \\ \hline
Nova             & 20,748                 & 14,680                                & 625                              \\ 
Neutron          & 13,541                 & 7,951                                 & 342                              \\ 
Qt Base          & 66,878                 & 35,021                                & 1,672                            \\ 
Qt Creator       & 26,015                 & 11,952                                & 574                              \\ \hline       
\textbf{Total}   & 127,182                & 69,604                                & 3,213                            \\ \hline
\end{tabular}
\end{table}

\subsubsection{Data Extraction and Analysis}
\textbf{1) Data Extraction}

Before data analysis, the first and second authors conducted a pilot data extraction by randomly selecting 10 code snippet comments from each of the four project in each year (80 review comments in total). The two authors extracted the data items listed in Table \ref{Data items extracted and their corresponding RQs} independently. If any disagreements arouse, the third author was involved to discuss with the two authors and came to an agreement. After the pilot data extraction, the first two authors extracted the data items from all the code snippet comments identified. During this process, any uncertain part was discussed between the first three authors until they reached a consensus to increase the correctness of the extracted data. Finally, the first author rechecked and compared all the extracted data to further enhance the accuracy of the extracted data.

\begin{table}[htbp]
\renewcommand\arraystretch{1.5}
\caption{Data items extracted and their corresponding RQs}
\label{Data items extracted and their corresponding RQs}
\begin{tabular}{m{0.5cm}<{\centering}m{2.4cm}m{7.3cm}m{0.5cm}<{\centering}}
\hline
\textbf{\#} & \textbf{Data Item}  & \textbf{Description}                                                                                                             & \textbf{RQ}   \\ \hline
D1          & Selected            & \textit{Whether or not a review comment contains code snippets}                                                                  & RQ1   \\ \hline
D2          & Identity            & \textit{The identity of the people who makes the review comments (i.e., reviewer or developer)}                                  & RQ1   \\ \hline
D3          & Purpose             & \textit{The intention of providing code snippets in a review comment}                                                            & RQ2   \\ \hline
D4          & Detailed purpose    & \textit{The elaborated intention of providing code snippets in a review comment}                                                 & RQ2   \\ \hline
D5          & Developer's action  & \textit{The action taken by the developer (i.e., accept, ignore, or not accept) towards the reviewer's code snippet suggestion}  & RQ3   \\ \hline
D6          & Evidence            & \textit{The proof that the developer accepted the code snippet suggestion}                                                       & RQ3   \\ \hline
D7          & Not accept reason   & \textit{The reason why the developer did not accept the reviewer's code snippet suggestion}                                      & RQ4   \\ \hline
\end{tabular}
\end{table}

Note that different data were used to answer the RQs. To answer RQ1, we used all the review comments after \textbf{step two} of data labelling (see Section \ref{datalabelling}) as the raw data, and we counted how many review comments in the raw data contain code snippets to investigate the extent of using code snippets in code reviews. To answer RQ2, we analyzed all the review comments that contain code snippets gained from RQ1 (i.e., the review comments after \textbf{step three}) to get the purposes of reviewers when using code snippets in code reviews. To answer RQ3, we collected the code snippet suggestions identified in RQ2 to explore the reaction of developers towards these suggestions, and we then used the unaccepted suggestions as the data to answer RQ4.

\textbf{2) Data Analysis}

After the data extraction, the first and second authors analyzed the extracted data to answer the RQs of the exploratory study.

\noindent\textbf{RQ1: To what extent are code snippets used in code reviews?}

To answer RQ1, we analyzed (1) the percentage of review comments that contain code snippets, (2) the percentages of code snippets provided in review comments by reviewers and developers, and (3) the percentages of developers and reviewers who had provided code snippets in review comments, to investigate the usage of code snippets in code reviews. By analyzing these three aspects, we could know the basic overview of how often reviewers and developers provide code snippets in code reviews, as well as the main identity of people who provide code snippets in code reviews.

\noindent\textbf{RQ2: What are the purposes of code snippets provided by reviewers in code reviews?}

To answer RQ2, we used Open Coding and Constant Comparison methods from Grounded Theory (GT) \citep{stol2016grounded} to find the purposes of reviewers providing code snippets in code reviews. As the purposes of reviewers providing code snippets in code reviews might be expected to be closely related to the purposes of shared links and comments in general, we considered both taxonomies proposed by \cite{wang2021understanding} and \cite{li2017automatic} as the base to build the taxonomy in this study. For every review comment containing code snippets provided by reviewers, we read through the textual content of the comment, its corresponding source code, and its contextual information to understand the reviewer's purpose of providing code snippets. The detailed steps of data analysis for answering RQ2 are the following:
\begin{enumerate}[itemindent=1.5em]
    \item The first two authors coded reviewers' purposes of providing code snippets in code review comments by highlighting the text sections related.
    \item The first author rechecked all the coding results to make sure that the extracted data of RQ2 were correctly coded.
    \item The first two authors grouped similar codes into categories. The process is iterative, in which the two coders went back and forth between the codes and categories to refine the taxonomy.
    \item The third author then examined the analysis results and disagreements were eliminated through discussions with the first two authors.
\end{enumerate}

A review comment may contain multiple code snippets, in which case we fully considered the contextual information and selected the most significant purpose as the final purpose of providing code snippets in this review comment. The reason why we did not identify multiple purposes in review comments with multiple code snippets is that a piece of code review comment is an integral part of delivering code review information, which could not be analyzed in separate sections. For example, a reviewer provided two pieces of code snippets in the following review comments. The first code snippet serves as a specific code use case to assist reviewers in elaborating on the aforementioned example, while the second one was made by the reviewer to suggest the developer to weaken the return type of the function. Based on the comprehensive analysis and feedback from the developer, we concluded that the first code snippet which elaborates the example at length is used to help the developer better understand the flaws in the current code and thus better understand the suggestions in the second code snippet. Therefore, we selected the purpose of providing the second code snippet by the reviewer as the main purpose of the code snippets provided in this review comment. Note that only 10 (0.31\%, 10/3213) review comments contain multiple code snippets, which is very rare.

\begin{qoutebox}{white}{}
    \textbf{Link:} \url{http://alturl.com/n9zbr}\\
    \textbf{Reviewer:} ``In case if value is an lvalue, T will be deduced as lvalue ref, for example: [code snippet]... which is not what we want. You should decay the return type, i.e.: [code snippet]...''\\
    \textbf{Developer:} ``Done.''
\end{qoutebox}

\vspace{12pt}

\noindent\textbf{RQ3: How do developers treat code snippet suggestions in code reviews?}

\begin{figure}[htbp]
	\centering
	\includegraphics[width=1.0\linewidth]{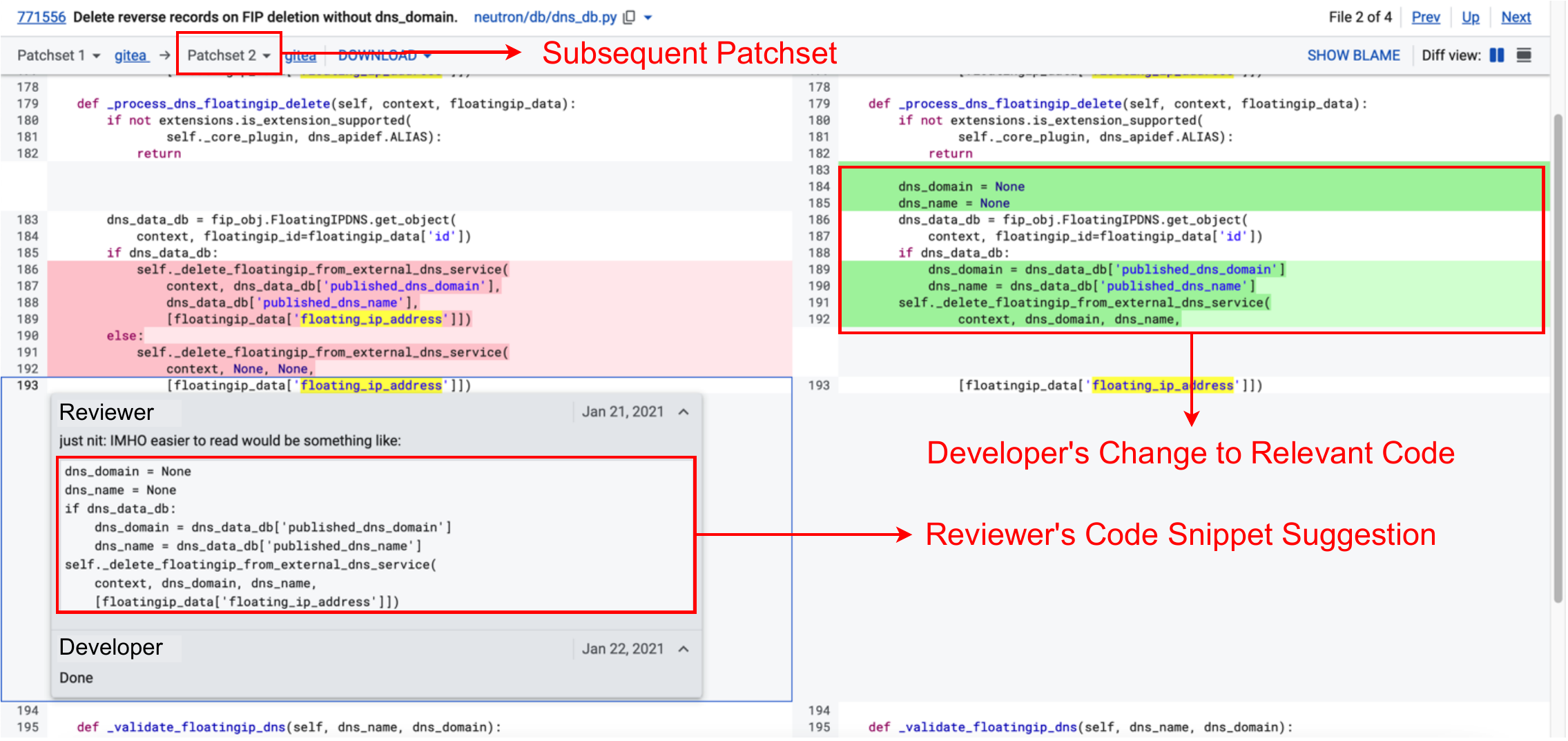}
	\caption{Example of a developer accepting a reviewer's code snippet suggestion}
	\label{fig: Example of a developer accepting a reviewer' code snippet suggestion}
\end{figure}

To answer RQ3, we manually checked and analyzed the contextual information of code snippet comments, including the whole code review discussions and associated source code. 
First, we looked through developers' responses to reviewers' code snippet suggestions. Second, we compared the suggested code snippets with the changes made by developers to the reviewed source code. We combined both textual and code information to conclude the acceptance of code snippet suggestions. According to our analysis results, the actions developers took towards reviewers' code snippet suggestions are in three categories:
\begin{itemize}
    \item \textbf{Accept:} (1) developers change the relevant code based on the code snippets provided by the reviewers or (2) developers clearly show a positive attitude toward the code snippet suggestions provided by the reviewers. \\
    For situation (1), Fig. \ref{fig: Example of a developer accepting a reviewer' code snippet suggestion} presents an example of a developer accepting a reviewer's code snippet suggestion. In this review comment, the reviewer provided a code snippet suggestion to make the code more readable (i.e., ``\textit{just nit: IMHO easier to read would be something like: [code snippet]}''), and the developer replied ``\textit{Done}'' and made corresponding code change to the relevant source code, which means that the developer accepted the code snippet suggestion.\\
    For situation (2), the reviewer provided a suggestion for optimizing the code and attached a relevant code snippet in the review comment below. The developer first thanked the reviewer and made it clear that he would try it out, which also indicates that the developer accepted the reviewer's code snippet suggestion.
    \begin{qoutebox}{white}{}
    \textbf{Link:} \url{http://alturl.com/hroeq}\\
    \textbf{Reviewer:} ``If we wanted to optimize as much as possible, we could limit to only the neutron\_pg\_drop row with something like: [code snippet]''\\
    \textbf{Developer:} ``Thanks for the tip, I'll definitely try that out...''
    \end{qoutebox}
    \item \textbf{Ignore:} developers neither respond to reviewers nor change the relevant code in the subsequent patchsets.
    \item \textbf{Not Accept:} (1) developers articulate totally different opinions towards reviewers' code snippet suggestions or (2) developers respond to reviewers but the developers do not change the relevant code or they do not change the relevant code according to the reviewers' suggestions.\\
    In the following review comment, the developer said that he would not change the current code to make the code more complex, which clearly shows that he would not accept the reviewer's code snippet suggestion.
    \begin{qoutebox}{white}{}
    \textbf{Link:} \url{http://alturl.com/vpw3e}\\
    \textbf{Reviewer:} ``How about adding: [code snippet] ...''\\
    \textbf{Developer:} ``... libvirt will do something similar but until we actually need that for some reason i would prefer not to add the extra complexity.''
    \end{qoutebox}
\end{itemize}

This process was performed by the same two coders in the data analysis of RQ2. Most of the identified code snippet suggestions in RQ2 are trivial cases which are straightforward and simple to tell whether they were accepted, ignored, or not accepted. Of all the 2,322 code snippet suggestions identified in OpenStack and Qt communities, the two coders could not decide on 28 of the suggestions. For the 28 code snippet suggestions, the third author was involved to discuss with the two coders and reach an agreement.



\noindent\textbf{RQ4: What are the reasons that developers do not accept code snippet suggestions in code reviews?}

To answer RQ4, we extracted the relevant data item listed in Table \ref{Data items extracted and their corresponding RQs} and analyzed all the code snippet suggestions which were not accepted by developers based on the results of RQ3. Through manually checking on the whole corresponding discussions among developers and reviewers, we used Open Coding and Constant Comparison methods to identify the reasons that developers do not accept code snippet suggestions in code reviews, and this process was similar to the data analysis for answering RQ2.

\subsection{Survey Study Design}
A survey is used to gather information from or about people to describe, compare, or explain their knowledge, attitudes, and behavior \citep{fink2003survey}. To validate the results of RQ1 and RQ2 and to obtain practitioners' insights on code snippets in code reviews (e.g., usage scenarios), we decided to conduct an industrial survey as a complementary data collection tool to understand practitioners' perceptions and current practices regarding code snippets in code reviews. A survey study could be conducted by self-administered questionnaires, telephone surveys, and one-to-one interviews to collect data \citep{kitchenham2008personal}. We finally chose to employ an online self-administered questionnaire to collect the information because using the online questionnaire could help us get evidence from potential participants which may come from different countries and facilitate the collection of responses from a large number of participants \citep{campbell2013coding}.

\subsubsection{Creating the Questionnaire and Recruitment of Participants}
We formulated the survey questionnaire covering five RQs (RQ1, RQ2, RQ5, RQ6, and RQ7). The questionnaire was prepared in both English and Chinese in order to get more responses. The questionnaire in English could help us invite potential participants from the World, and the questionnaire in Chinese could increase the probability of practitioners in China filling them out. To ensure that the meaning of the questionnaire is the same in both languages, the first two authors who are native Chinese speakers translated the questionnaire in English into Chinese, and the third author checked and refined the translation. The questionnaire is composed of seven parts: the Welcome page shows the questionnaire requirements, introduction, and an example of code snippet used in review comments, one question (SQ1) about participants' background information, two questions (SQ2 and SQ3) about validating RQ1 and RQ2, and three questions (SQ4, SQ5, and SQ6) about answering RQ5, RQ6, and RQ7 (see Table \ref{Survey questions on code snippets in code reviews} and Table \ref{Survey questions and their analysis methods for answering the RQs}). After preparing the survey questionnaire, we needed to select the survey participants. Our target participants are software developers with code review experience, and we used the following contact channels to invite the potential respondents with code review experience:
\begin{itemize}
    \item \textbf{Developers from the OpenStack and Qt communities:} To engage developers from the OpenStack and Qt communities in our survey study, we employed two ways: (1) send survey invitation emails to the 640 developers who had provided code snippets in code reviews collected in the dataset of our exploratory study and (2) send a survey invitation email to the developer mailing lists of OpenStack and Qt. For (1), when sending the survey invitation emails to the 640 developers, we first apologized for bothering them. Besides, we only emailed these developers once. For (2), we also sent the survey invitation email only one time to the developer mailing lists of OpenStack and Qt in order to contact a wide range developers of OpenStack and Qt. With these measures, developers from OpenStack and Qt could voluntarily choose whether to fill out the survey questionnaire with minimal interruption.
    \item \textbf{Developers from well-known software companies:} We contacted developers from well-known software companies in China, i.e., ByteDance, Alibaba, and Tencent, to participate in our survey study. These companies are the leading and most prestigious IT companies in China. ByteDance and Alibaba have been credited as the top 50 most innovative companies in 2023\footnote{\url{http://alturl.com/wef4g}}, while Tencent is the world's tenth most valuable company as of Feburary 2022\footnote{\url{http://alturl.com/n9rgs}}. To collect as many responses as possible, we also employed snowball sampling \citep{shull2007guide} by requesting theses developers to share the survey invitation with individuals or groups deemed as potential participants.
    \item \textbf{Developers from professional software development groups:} We posted the survey invitation on software development groups listed in Table \ref{Posted Linkedin groups and the corresponding post links} on Linkedin, where software developers from around the world share their issues, experiences, and knowledge.
    \begin{table}[htbp]
    \renewcommand\arraystretch{1.5}
    \caption{Software development groups in Linkedin used to post our survey}
    \label{Posted Linkedin groups and the corresponding post links}
    \begin{tabular}{m{1cm}<{\centering}m{6.5cm}m{3.5cm}}  \hline
    \textbf{\#}     & \textbf{Linkedin Group}                                                                           & \textbf{URL}                 \\ \hline
    \textbf{LP1}    & Software Developer                                                                                & \url{http://alturl.com/aybty} \\ \hline
    \textbf{LG2}    & Agile and Lean Software Development                                                               & \url{http://alturl.com/fgvm3} \\ \hline
    \textbf{LP3}    & Software Engineer - Full Stack Developer                                                          & \url{http://alturl.com/cxkxo} \\ \hline
    \textbf{LP4}    & Java / J2EE / Core Java / Corejava / Java Developer / Software Engineer - (JAVA)                  & \url{http://alturl.com/bzd3y} \\ \hline
    \textbf{LP5}    & Python Developer / Full Stack Developer / Web Developer / Software Developer / Data Analyst       & \url{http://alturl.com/uoqz5} \\ \hline
    \textbf{LP6}    & Software Developer, Programmer and Architect ( Java | Python | PHP | C\# | C++ | GO | Swift)      & \url{http://alturl.com/v7y5p} \\ \hline
    \textbf{LP7}    & Software Engineer / Developer / Programmer / Data Analyst / Data Scientist / Data Engineer / RPA  & \url{http://alturl.com/5h74j} \\ \hline
    \end{tabular}
    \end{table}
\end{itemize}

Note that throughout the entire survey study, we upheld the privacy of OpenStack and Qt community members while gathering and utilizing information regarding code reviews. We tried our best to make sure that our research was conducted ethically, ensuring the confidentiality of the data collected, and all the data acquired was solely utilized for research purposes.

\subsubsection{Evaluating and Validating the Questionnaire}
Before disseminating the survey invitations, we conducted a pilot survey with developers from the OpenStack and Qt communities to evaluate and validate the questionnaire. For the 640 developers who had provided code snippets in code reviews that we collected in the dataset of our exploratory study, we ranked them based on the number of review comments they made as an indicator of activeness, and sent the survey invitations to the top 100 most active code reviewers.
Out of the 100 participants contacted for the pilot survey, 10 replied. By looking through the pilot results, we checked the understandability of the survey questions and the effectiveness of each question. We found that the length of the survey is appropriate, the questions are clear and easy to understand, and the answers to the questions are meaningful, and consequently we did not refine the questionnaire. Finally, our survey questionnaire\footnote{\url{https://forms.gle/7eKfjhzHtnBXhEMJ7}} is composed of 4 closed-ended questions and 2 open-ended questions. Table \ref{The welcome page of the survey} presents the Welcome page of the survey questionnaire, and Table \ref{Survey questions on code snippets in code reviews} presents the survey questions of the questionnaire.

\begin{table}[htbp]
\renewcommand\arraystretch{1.5}
\caption{The Welcome page of the survey questionnaire}
\label{The welcome page of the survey}
\centering
\begin{tabular}{|p{11.7cm}|}  
\hline  
\vspace{4pt}
\textbf{The goal} of this survey is trying to understand the practice and purposes of providing code snippets in code reviews. This questionnaire is designed to capture the knowledge and experience of industrial practitioners in the use of code snippets in code reviews. Our questionnaire contains 4 closed-ended questions and 2 open-ended questions, which may take about \textbf{3-5 minutes}. \\ 
\vspace{3pt}
Note that no personal information will be involved in this questionnaire and nor will your response be disclosed to the third parties. Please feel free to contact us if you have any questions or concerns.\\
\vspace{3pt}
Thank you very much for your participation! \\ 
\vspace{4pt}
\\ \hline
\vspace{4pt}
As shown in the example below, a reviewer made a suggestion towards code style and provided a \textbf{code snippet} to the developer in his/her code review comment, which intends to help the developer write more readable code. \\ 
\vspace{4pt}
\includegraphics[width=1.0\linewidth]{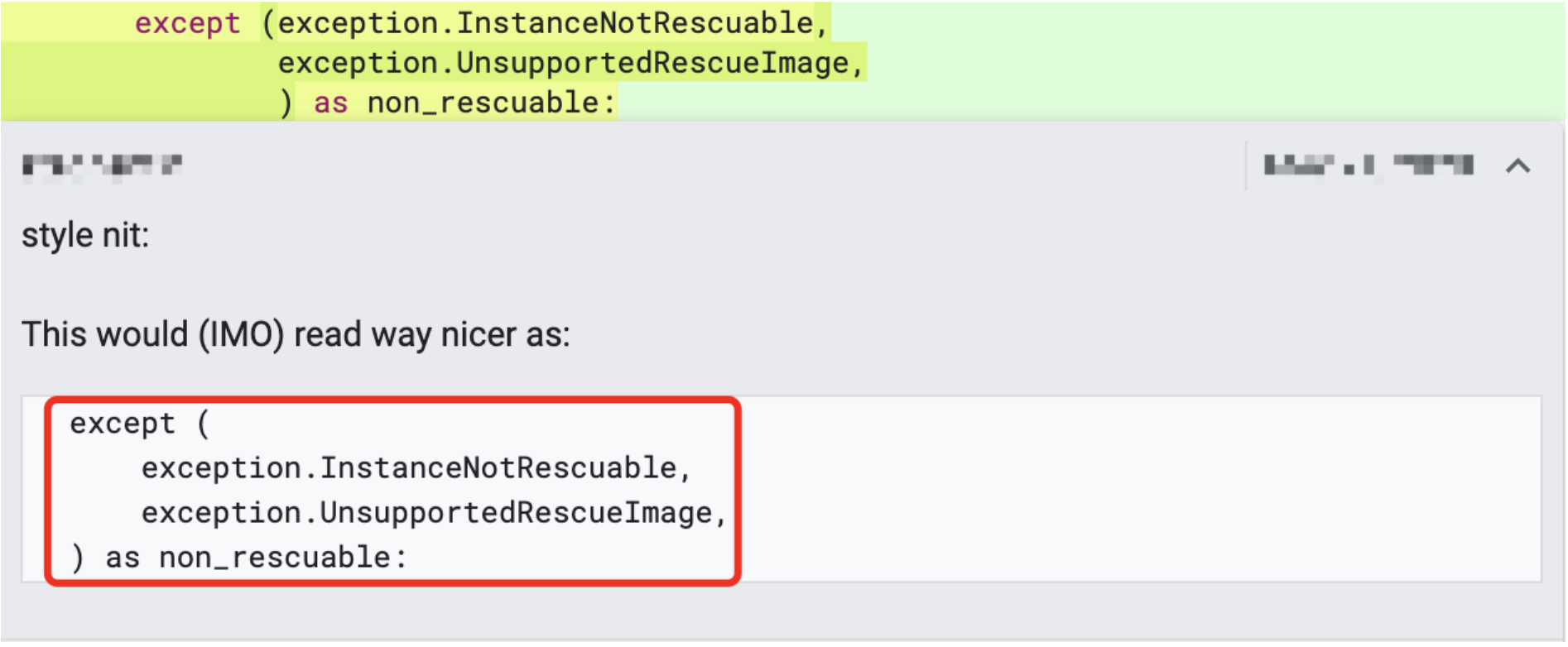}
\vspace{4pt} 
\\ \hline
\end{tabular}
\end{table}

\begin{table}[htbp]
\renewcommand\arraystretch{1.5}
\caption{Survey questions on code snippets in code reviews}
\label{Survey questions on code snippets in code reviews}
\centering
\begin{tabular}{|p{0.5cm}<{\centering}|p{2cm}|p{4cm}|p{4cm}|}  
\hline  
\textbf{ID}     &\textbf{Type of Questions}                   & \textbf{Questions}                        & \textbf{Type of Answers}  \\ \hline
SQ1 &Background information about participants    & Q1. How many years have you been involved in software development?                              
& \textless 1 year / 1 $\sim$ 3 years / 3 $\sim$ 5 years / \textgreater 5 years \\ \hline
SQ2 &Question for validating RQ1                  & Q2. How often do you provide code snippets in review comments when conducting code reviews?     
& Never / Very few / Sometimes / Often \\ \hline
SQ3 &Question for validating RQ2                  & Q3. (Multiple Choice) As a reviewer in code reviews, for what purposes do you provide code snippets in review comments?                           
& Improving Code Implementation - Point out alternative solutions or advice to improve the current code in the patchsets (e.g., design or detailed implementation). / Following Code Style - Make the style of the current code consistent with the best code conventions. / Correcting Code - Show what kind of mistakes developers have made in current code and to correct the error code. / Complementing Code Implementation - Remind developers that the current code implementations are incomplete and they should complement the code with the provided code snippets. / Elaborating - Help reviewers supplement their explanations or illustrations. / Providing Context - Provide additional information related to what reviewers had said in the review comments. / Other \\ \hline
SQ4 &Question for answering RQ5                   & Q4. (Optional) As a reviewer in code reviews, in which scenarios you tend to provide code snippets in code review comments?     
& Free text \\ \hline
SQ5 &Question for answering RQ6                   & Q5. As a developer in code reviews, what is your attitude towards the provided code snippets in code reviews?                  
& Positive. I think it is a good thing to see review comments that contain code snippets. / Neutral. I don't care if review comments contain code snippets or not. / Negative. I think it is troublesome to see review comments that contain code snippets. / Other  \\ \hline
SQ6 &Question for answering RQ7                   & Q6. (Optional) As a developer in code reviews, what characteristics of code snippets would you like reviewers to provide in code review comments?   & Free text \\ \hline
\end{tabular}
\end{table}

\subsubsection{Conducting the Survey and Analyzing Survey Data}
After finalizing the questionnaire, we sent out the survey invitations to the participants we recruited. The invitations were sent on March 26, 2023, and as of May 31, 2023, a total of 63 responses were collected. For the 63 participants who filled out the survey questionnaire, 43 are developers from OpenStack and Qt communities or LinkedIn groups, while 17 are developers from well-known software companies in China. However, we were unable to tell the response rate for both open-source and closed-source cases. This is because we lacked the information regarding the number of developers who received our survey invitation emails from the developer mailing lists of OpenStack and Qt. Additionally, the anonymity of the survey responses made it challenging to identify the sources of the respondents.

We used descriptive statistics \citep{kaur2018descriptive}, Open Coding, and Constant Comparison techniques \citep{glaser2017discovery} to analyze the quantitative (i.e., closed-ended questions) and qualitative (i.e., open-ended questions) responses to the survey questions. Table \ref{Survey questions and their analysis methods for answering the RQs} presents the survey questions and their analysis methods for answering the RQs. 

\begin{table}[htbp]
\renewcommand\arraystretch{1.5}
\centering
\caption{Survey questions and their analysis methods for answering the RQs}
\label{Survey questions and their analysis methods for answering the RQs}
\begin{tabular}{p{3cm}<{\centering}lp{2cm}<{\centering}}
\hline
\textbf{Survey Question} & \textbf{Data Analysis Method}           & \textbf{RQ}           \\ \hline
SQ1                       & Descriptive Statistics                 & Demographic           \\ \hline
SQ2                       & Descriptive Statistics                 & RQ1                   \\ \hline
SQ3                       & Descriptive Statistics                 & RQ2                   \\ \hline
SQ4                       & Open Coding and Constant Comparison    & RQ5                   \\ \hline
SQ5                       & Descriptive Statistics                 & RQ6                   \\ \hline
SQ6                       & Open Coding and Constant Comparison    & RQ7                   \\ \hline
\end{tabular}
\end{table}

\begin{figure}[htbp]
	\centering
	\includegraphics[width=0.85\linewidth]{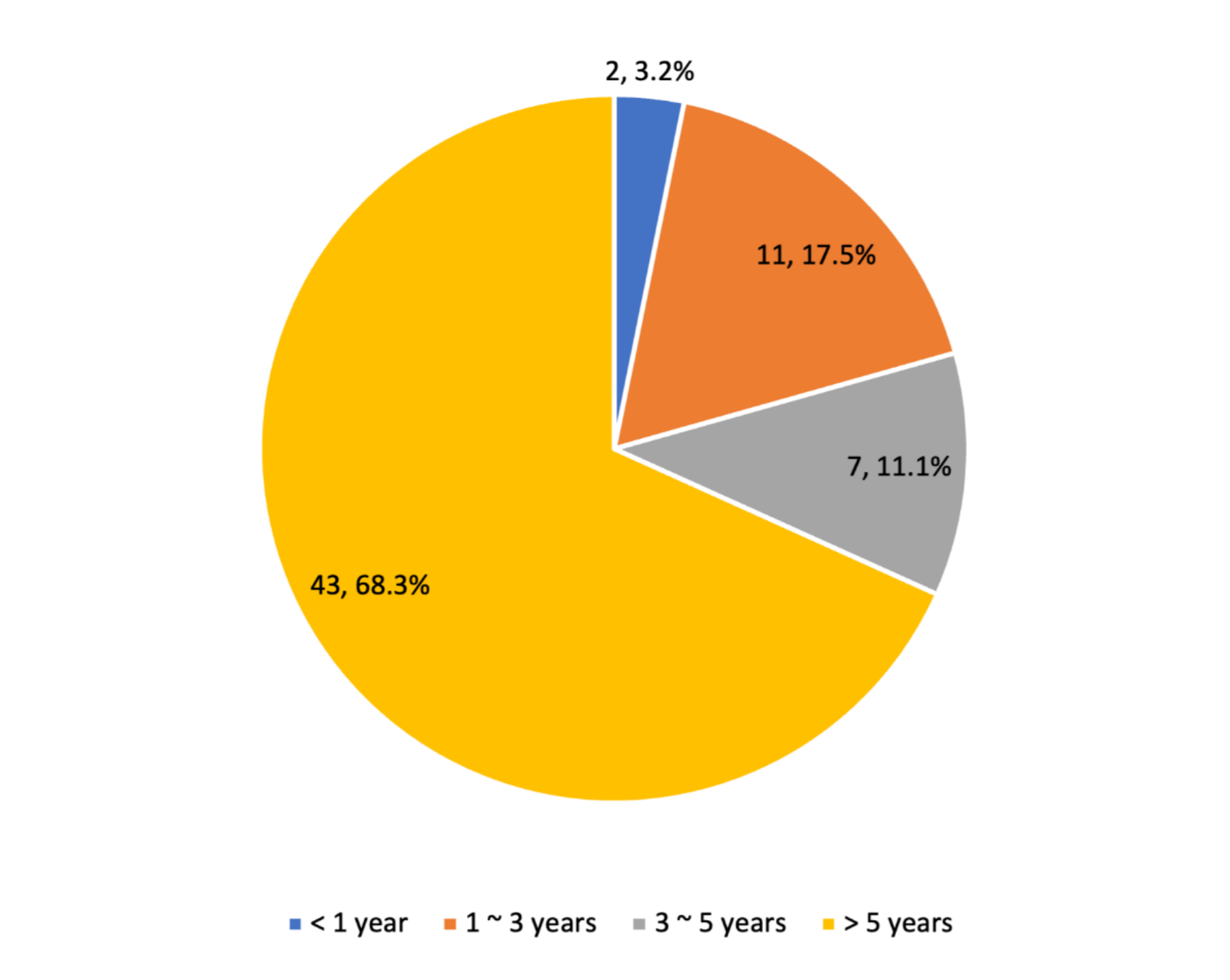}
	\caption{Distribution of software development experience of the participants}
	\label{fig: Distribution of software development work experience of the participants}
\end{figure}

According to the survey results, the distribution of software development experience of the survey participants is shown in Fig.~\ref{fig: Distribution of software development work experience of the participants}. We can find that 79.4\% (11.1\% + 68.3\% = 79.4\%) of the participants have more than 3 years of software development experience, and nearly 70\% of the participants have more than 5 years of software development experience, which somewhat indicates that the survey results are representative.

Note that two examples are provided for the two open-ended survey questions respectively to help participants better understand the terms used in the questions. The examples were formed based on the results of the exploratory study, which may somewhat restrict practitioners from answering the two open-ended survey questions. In order to get valid survey results, we excluded the responses that are inconsistent with the question, randomly filled in, or meaningless when we analyzed the data. During the process of analyzing the data, the first three authors discussed inconsistent opinions till they reached an agreement. The survey results have been provided in our dataset \citep{replpack}.

\section{Results}
\label{sec:results}
In this section, we present the final results of the seven RQs formulated in Section \ref{subsec:research_questions}. 

\subsection{RQ1: The Extent Code Snippets Used in Code Reviews}
\label{subsec:results_of_RQ1}

\subsubsection{Results of RQ1}
To answer RQ1, we investigated (1) the proportion of review comments that contain code snippets, (2) the proportions of code snippets provided in review comments by reviewers and developers respectively, and (3) the proportion of developers and reviewers who had provided code snippets in review comments.

\vspace{10pt}

\noindent\textbf{The proportion of review comments that contain code snippets}

Table \ref{Counts and Percentages of Review Comments that Contain Code Snippets per Community} presents the counts and percentages of review comments that contain code snippets per community. In total, we collected 69,604 review comments from OpenStack and Qt, 3,213 of which contain code snippets. Overall, the percentage of code snippet comments is 4.6\% only, which means that code snippets are not prevalently used in review comments by during code review. Moreover, the percentages of code snippet comments are similar in OpenStack and Qt, thus indicating a consistency of the results.

\begin{table}[htbp]
\renewcommand\arraystretch{1.5}
\centering
\caption{Counts and percentages of review comments that contain code snippets}
\label{Counts and Percentages of Review Comments that Contain Code Snippets per Community}
\begin{tabular}{lm{1.2cm}<{\centering}m{6cm}<{\centering}}
\hline
\textbf{Community} & \textbf{\#RCs} & \textbf{\#RCs with Code Snippets (CSs) (\%)} \\ \hline
OpenStack          & 22,631                   & 967 (4.3\%)                                     \\ 
Qt                 & 46,973                   & 2,246 (4.8\%)                                    \\ \hline       
\textbf{Total}     & 69,604                   & 3,213 (4.6\%)                                    \\ \hline
\end{tabular}
\end{table}

\begin{table}[htbp]
\renewcommand\arraystretch{1.5}
\centering
\caption{Counts and percentages of review comments that contain code snippets provided by reviewers and developers}
\label{Counts and Percentages of Review Comments that Contain Code Snippets Provided by Reviewers and Developers}
\begin{tabular}{lm{2.8cm}<{\centering}m{2.8cm}<{\centering}}
\hline
\textbf{Community} & \textbf{\#RCs with CSs by Reviewers (\%)}   & \textbf{\#RCs with CSs by Developers (\%)} \\ \hline
OpenStack          & 863 (89.2\%)                                & 104 (10.8\%)                       \\
Qt                 & 1,905 (84.8\%)                              & 341 (15.2\%)                       \\ \hline
\textbf{Total}     & 2,768 (86.2\%)                              & 445 (13.8\%)                       \\ \hline
\end{tabular}
\end{table}

\vspace{10pt}

\noindent\textbf{The proportion of code snippets provided in review comments by reviewers and developers respectively}

Table \ref{Counts and Percentages of Review Comments that Contain Code Snippets Provided by Reviewers and Developers} shows the counts and percentages of code snippet comments provided by reviewers and developers respectively. In code reviews, most review comments that contain code snippets are provided by reviewers, whose proportions in OpenStack and Qt are 89.2\% and 84.4\% respectively. In general, more than 85\% (2,768 out of 3,213, 86.2\%) review comments with code snippets are provided by reviewers. It is not surprising that reviewers are the primary source that provides code snippet comments because reviewers play a major role in code reviews. 

\vspace{10pt}

\noindent\textbf{The proportion of developers and reviewers who had provided code snippets in review comments}

Table \ref{Counts and Percentages of developers and reviewers who had provided code snippets in code reviews per community} presents the proportion of developers and reviewers who had provided code snippets in review comments according to the dataset we collected from OpenStack and Qt. From the results, we can find that 25.4\% developers and reviewers in OpenStack had provided code snippets, while 34.4\% in Qt. Overall, the proportion of developers and reviewers who had provided code snippets in review comments is 30.2\%. However, given the reason that in a project a contributor can play both roles, i.e., developer and reviewer, we could not tell the proportion of developers or reviewers providing code snippets in review comments.

\begin{table}[htbp]
\centering
\renewcommand\arraystretch{1.5}
\caption{Counts and Percentages of developers and reviewers who had provided code snippets in code reviews}
\label{Counts and Percentages of developers and reviewers who had provided code snippets in code reviews per community}
\begin{tabular}{lm{3cm}<{\centering}m{4.5cm}<{\centering}c}
\hline
\textbf{Community} & \textbf{\#Developers and Reviewers} & \textbf{\#Developers and Reviewers who had Provided CSs} & \% \\ \hline
OpenStack          & 338                                 & 86                                                       & 25.4\%         \\
Qt                 & 375                                 & 129                                                      & 34.4\%        \\ \hline
\textbf{Total}     & 713                                 & 215                                                      & 30.2\%       \\ \hline
\end{tabular}
\end{table}

\subsubsection{Feedback on the Findings of RQ1}
\begin{figure}[htbp]
	\centering
	\includegraphics[width=1.0\linewidth]{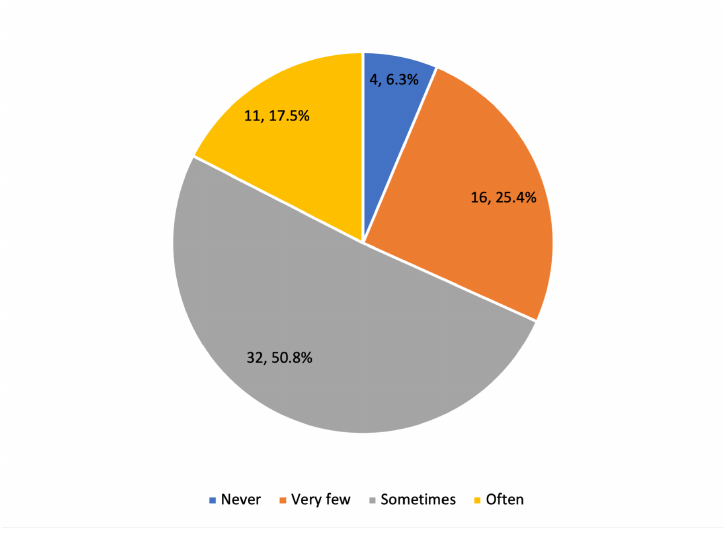}
	\caption{Distribution of how often industrial developers provide code snippets in code reviews}
	\label{fig: Feedback_on_Findings_of_RQ1}
\end{figure}

Fig.~\ref{fig: Feedback_on_Findings_of_RQ1} shows the distribution of how often industrial developers provide code snippets in review comments when conducting code reviews. We can find that most participants said that they provided code snippets in code reviews sometimes or very few, accounting for 76.2\% in total. More specifically, 50.8\% participants chose that they sometimes used code snippet in code reviews and 25.4\% chose that they used code snippets very few. 17.5\% participants said that they often used code snippets in code reviews. Only 4 participants expressed that they never used code snippets when conducting code reviews, accounting for 6.3\%. According to the feedback, most developers tend to use code snippets in code review process, but how often they use code snippets mainly depends on the necessity of the context (reviewed code and review discussions) rather than the perceived value of using code snippets in review comments.

\noindent\begin{center}
		\begin{tcolorbox}[colback=black!5, colframe=black!20, width=1.0\linewidth, arc=1mm, auto outer arc, boxrule=1.5pt]                       
            {{\textbf{RQ1 Summary:} According to our results, only a small proportion of review comments contain code snippets (less than 5\%), and most code snippet comments are provided by reviewers rather than developers (more than 85\%).}}
		\end{tcolorbox}
\end{center}

\subsection{RQ2: The Reviewers' Purposes of Providing Code Snippets in Code Reviews}
\label{subsec:results_of_RQ2}

\subsubsection{Results of RQ2}
\label{subsubsec: results_of_RQ2}
\label{sub_results_of_rq2}

\noindent\textbf{The purposes of reviewers providing code snippets in code reviews}

\begin{figure}[htbp]
	\centering
	\includegraphics[width=1.0\linewidth]{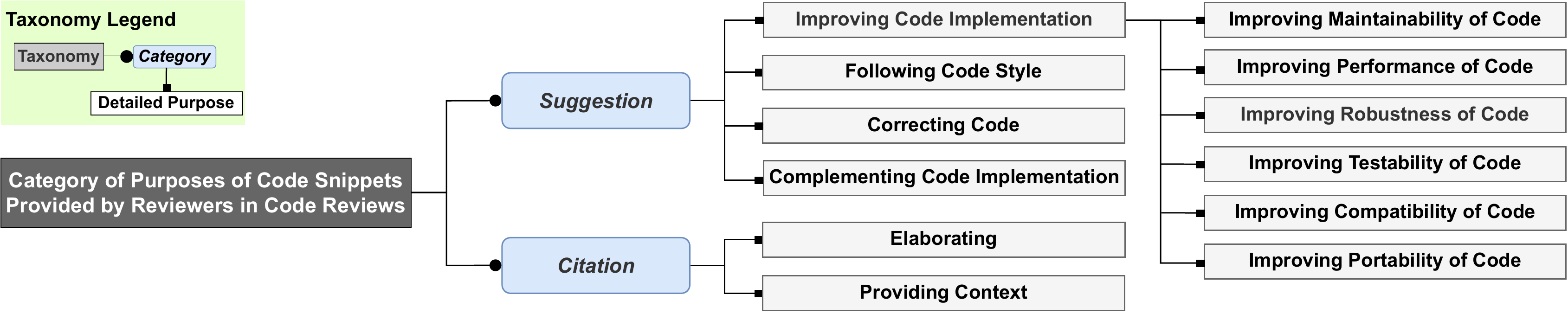}
	\caption{Category of purposes of code snippets provided by reviewers in code reviews}
	\label{fig: Category of purposes of code snippets provided by reviewers in code reviews}
\end{figure}

To answer RQ2, we used Open Coding and Constant Comparison methods to identify the purposes of reviewers providing code snippets in code reviews, and two high-level categories are formulated (i.e., \textit{Suggestion} and \textit{Citation}). We further identified four detailed purposes of providing code snippets in code reviews under the \textit{Suggestion} category and two detailed purposes under the \textit{Citation} category. Moreover, to make the detailed purpose \textit{Improving Code Implementation} clearer, six specific purposes are categorized under it (see Fig.~\ref{fig: Category of purposes of code snippets provided by reviewers in code reviews}). The identified purposes are presented below, and we provide a review comment example for each detailed purpose.

\textbf{(1) Suggestion} refers to the situation in which reviewers provide code snippets in review comments to recommend developers what they can do or what they should do to make the quality of code better. The provided code snippets are what reviewers want developers to change the current code to. \textit{Suggestion} contains four detailed purposes:

\textbf{Improving Code Implementation:} The reviewed code is correct and complete but needs improvements. The code snippets provided by reviewers in code reviews are used to remind developers what they can do to improve specific quality attributes of current code:
\begin{itemize}
    \item \textbf{Improving Maintainability of Code:} The provided code snippets are what reviewers want developers to change the current code to, which can make code simpler, more readable, more efficient, and thus easier to maintain.
    \begin{qoutebox}{white}{}
    \textbf{Link:} \url{http://alturl.com/am3q9} \\
    \textbf{Reviewer:} ``Sure, I'm just saying we're supposing that check\_traits will always return this way. I'm just saying (for code maintenance) that having some explicit attributes would help (and would raise some error in case the returned object changes). eg. [code snippet]''
    \end{qoutebox}
    \item \textbf{Improving Performance of Code:} The provided code snippets are what reviewers want developers to change the current code to, which can make code execute faster and be more efficient.
    \begin{qoutebox}{white}{}
    \textbf{Link:} \url{http://alturl.com/4gcs7} \\
    \textbf{Reviewer:} ``I'll let Nate comment on this too, but the problem now is there will always be two lookups for the SG now, which could affect performance. One way to lessen this would be a slight change here, something like: [code snippet]...''
    \end{qoutebox}
    \item \textbf{Improving Robustness of Code:} The provided code snippets are what reviewers want developers to change the current code to, which can make code have stronger ability to withstand or overcome adverse conditions.
    \begin{qoutebox}{white}{}
    \textbf{Link:} \url{http://alturl.com/uwr3h} \\
    \textbf{Reviewer:} ``This should not be computed by len / oldLength, which may overflow for small oldLength and large len. We know oldLength is of comparable order to dx() and dy(), so [code snippet] is more numerically robust.''
    \end{qoutebox}
    \item \textbf{Improving Testability of Code:} The provided code snippets are what reviewers want developers to change the current code to, which can make code support testing to a greater degree in a given test environment.
    \begin{qoutebox}{white}{}
    \textbf{Link:} \url{http://alturl.com/rfz6y} \\
    \textbf{Reviewer:} ``... i guess this is why you said almost all places in the commit message.maybe you should still use the fixture here too for consitency so add a setup thatdoes [code snippet]. this works but it might be nice have this bevhior in all tests by defualt.''
    \end{qoutebox}
    \item \textbf{Improving Compatibility of Code:} The provided code snippets are what reviewers want developers to change the current code to, which can make code enhance the ability to work with data or configurations created by older versions. 
    \begin{qoutebox}{white}{}
    \textbf{Link:} \url{http://alturl.com/oq69p} \\
    \textbf{Reviewer:} ``Why even bother with the \_WITH\_ARGS() version? We have macro varargs now. At the very least, make life easy for clients: [code snippet]. We could, of course, do similar for Q\_GLOBAL\_STATIC(), although we'd have to keep the \_WITH\_ARGS() version for backwards compatibility.''
    \end{qoutebox}
    \item \textbf{Improving Portability of Code:} The provided code snippets are what reviewers want developers to change the current code to, which can make code run on a wider variety of platforms.
    \begin{qoutebox}{white}{}
    \textbf{Link:} \url{http://alturl.com/urdmh} \\
    \textbf{Reviewer:} ``Would [code snippet] work, too? If so I'd prefer that as this doesn't depend directly on the host os and potentially works remotely.''
    \end{qoutebox}
\end{itemize}

\textbf{Following Code Style:} The reviewed code is correct and complete but has code style issues. The code snippets provided by reviewers in code reviews are used to remind developers the code style issues of current code, and developers should change the current code style to the style of provided code snippets, to be consistent with the best code conventions.
    \begin{qoutebox}{white}{}
    \textbf{Link:} \url{http://alturl.com/jis7g} \\
    \textbf{Reviewer:} ``I think this is not an acceptable coding style for Qt in this case. Would go with something like: [code snippet]...''
    \end{qoutebox}
    
\textbf{Correcting Code:} The reviewed code is incorrect (e.g., has logic errors). The code snippets provided by reviewers in code reviews are used to remind developers the incorrectness of current code, and developers should correct the error code with the provided code snippets.
    \begin{qoutebox}{white}{}
    \textbf{Link:} \url{http://alturl.com/4c63s} \\
    \textbf{Reviewer:} ``Hi, This is a BUG. Should modify it like this: [code snippet]...''\\
    \textbf{Developer:} ``Thanks much for pointing it out! I will investigate the issue.''
    \end{qoutebox}

\textbf{Complementing Code Implementation:} The reviewed code is incomplete. The code snippets provided by reviewers in code reviews are used to remind developers the incompleteness of current code, and developers should complement the code with the provided code snippets. 
    \begin{qoutebox}{white}{}
    \textbf{Link:} \url{http://alturl.com/qt62k} \\
    \textbf{Reviewer:} ``The case of empty headers (or CRLF-only headers) would have size() == 0 but succeed. It’s probably worth adding. Suggestion: [code snippet].''\\
    \textbf{Developer:} ``Done.''
    \end{qoutebox}

\vspace{10pt}

\textbf{(2) Citation} refers to the situation in which reviewers provide code snippets cited from internal (e.g., the code in the current files of the patchsets) or external (e.g., the code in other files of the project) sources in code reviews to offer developers necessary review-related information rather than offer suggestions. \textit{Citation} contains two detailed purposes:

\textbf{Elaborating:} The code snippets cited by reviewers in code reviews are used to help reviewers supplement their explanations or illustrations.
\begin{qoutebox}{white}{}
\textbf{Link:} \url{http://alturl.com/3h8iq} \\
\textbf{Reviewer:} ``I think you’re just getting confused because we’re building the regex in two steps. metakey\_pattern\_base isn’t being sent to the regex engine - we’re using it to compose the regex that is. Maybe this helps: [code snippet].''
\end{qoutebox}
When reviewers provide code snippets in code reviews for the purpose of \textit{Elaborating}, they usually have offered suggestions in the previous comments and they cite code snippets in a following new comment to explain why they make the previous suggestions.

\textbf{Providing Context:} The code snippets cited by reviewers in code reviews are used to provide contextual information related to what reviewers had mentioned in the review comments. 
\begin{qoutebox}{white}{}
\textbf{Link:} \url{http://alturl.com/8rzjs} \\
\textbf{Reviewer:} ``In `Ml2Plugin.\_create\_port\_db' the variable passed is the dictionary: [code snippet]''
\end{qoutebox}
When reviewers provide code snippets in code reviews for the purpose of \textit{Providing Context}, they usually cite code snippets from other files of the same project to make developers get the contextual information. Reviewers may also provide links related to the provided code snippets at the same time. 

\vspace{10pt}

\noindent\textbf{The proportion of code snippets provided in review comments for different purposes by reviewers}

We further investigated the distribution of the purposes of code snippets provided by reviewers in code reviews. Table \ref{The distribution of high-level purposes of using code snippets in the OpenStack and Qt communities} presents the distribution of the two high-level purposes of providing code snippets in the OpenStack and Qt communities. In general, among the 2,768 review comments with code snippets provided by reviewers, 2,322 (83.9\%) are provided for the purpose of \textit{Suggestion}, while 446 (16.1\%) for the purpose of \textit{Citation}. In both OpenStack and Qt communities, the main purpose of reviewers using code snippets in code reviews is \textit{Suggestion}, and the proportion of review comments for \textit{Suggestion} purpose is significantly higher than the proportion of \textit{Citation}. This result shows that code snippets are usually provided by reviewers in review comments to make suggestions for developers.

Moreover, Table \ref{The distribution of each detailed purpose of using code snippets in code reviews} presents the distribution of the detailed purposes under the two high-level purposes of using code snippets in code reviews. \textit{Improving Code Implementation} is the main purpose of reviewers providing code snippets in code reviews, which is more than 50\% in both OpenStack and Qt, and higher than other purposes by a large margin, indicating that reviewers often provide code snippets to offer developers suggestions which can improve quality attributes of current code. Among all the quality attributes, reviewers want to improve the maintainability of reviewed code most. Besides, the proportions of \textit{Correcting Code}, \textit{Complementing Code Implementation}, \textit{Elaborating}, and \textit{Providing Context} are very close in both communities. However, the proportion of \textit{Following Code Style} varies considerably between the two communities. In OpenStack, \textit{Following Code Style} accounts for 16.6\%, making it the second most frequent purpose of providing code snippets by reviewers in code reviews. However in Qt, \textit{Following Code Style} accounts for only 4.8\%.

\begin{table}[htbp]
\centering
\renewcommand\arraystretch{1.5}
\caption{Counts and percentages of review comments for high-level purposes of using code snippets in the OpenStack and Qt communities}
\label{The distribution of high-level purposes of using code snippets in the OpenStack and Qt communities}
\begin{tabular}{lm{2.8cm}<{\centering}m{3cm}<{\centering}m{3cm}<{\centering}}
\hline
\textbf{Community} & \textbf{\#RCs with CSs by Reviewers}  & \textbf{\#RCs with Suggestion Purpose}  & \textbf{\#RCs with Citation Purpose}\\ \hline
OpenStack          & 863                                                   & 747 (86.6\%)                                               & 116 (13.4\%)                              \\
Qt                 & 1,905                                                 & 1,575 (82.7\%)                                             & 330 (17.3\%)                              \\ \hline
\textbf{Total}     & 2,768                                                 & 2,322 (83.9\%)                                             & 446 (16.1\%)                              \\ \hline
\end{tabular}
\end{table}

\begin{table}[htbp]
\renewcommand\arraystretch{1.5}
\caption{Counts and percentages of review comments for each detailed purpose of using code snippets in code reviews}
\label{The distribution of each detailed purpose of using code snippets in code reviews}
\begin{tabular}{llcc}\hline
\multicolumn{2}{l}{\textbf{Detailed Purpose}}           & \textbf{\begin{tabular}[c]{@{}c@{}}\#RCs in\\OpenStack (\%)\end{tabular}} & \textbf{\begin{tabular}[c]{@{}c@{}}\#RCs in\\Qt (\%)\end{tabular}} \\ \hline
\multicolumn{1}{l}{\multirow{7}{*}{\begin{tabular}[c]{@{}l@{}}Improving Code \\ Implementation\end{tabular}}}                                          
& Improving Maintainability of Code                     & 416 (48.2\%)                                                              & 1,052 (55.2\%) \\
\multicolumn{1}{l}{} & Improving Performance of Code    & 11 (1.3\%)                                                                & 33 (1.7\%) \\
\multicolumn{1}{l}{} & Improving Robustness of Code     & 4 (0.5\%)                                                                 & 20 (1.0\%) \\
\multicolumn{1}{l}{} & Improving Testability of Code    & 1 (0.1\%)                                                                 & 6 (0.3\%) \\
\multicolumn{1}{l}{} & Improving Compatibility of Code  & 0 (0.0\%)                                                                 & 2 (0.1\%) \\
\multicolumn{1}{l}{} & Improving Portability of Code    & 0 (0.0\%)                                                                 & 1 (0.1\%) \\
\multicolumn{2}{l}{Following Code Style}                & 143 (16.6\%)                                                              & 91 (4.8\%) \\
\multicolumn{2}{l}{Correcting Code}                     & 90 (10.4\%)                                                               & 195 (10.2\%) \\
\multicolumn{2}{l}{Complementing Code Implementation}   & 82 (9.5\%)                                                                & 175 (9.2\%) \\ \hline
\multicolumn{2}{l}{Elaborating}                         & 72 (8.3\%)                                                                & 217 (11.4\%) \\
\multicolumn{2}{l}{Providing Context}                   & 44 (5.1\%)                                                                & 113 (5.9\%) \\ \hline
\multicolumn{2}{l}{\textbf{Total}}                      & 863                                                                       & 1,905 \\ \hline
\end{tabular}
\end{table}

\vspace{10pt}

\noindent\textbf{The proportion of reviewers who had provided code snippet comments with different purposes}

In total, there are 86 reviewers who had provided code snippets in OpenStack, and 101 in Qt. We then counted the reviewers who provided code snippet comments for different purposes, as shown in Table \ref{Counts and percentages of reviewers who had provided code snippets in code reviews with different purposes}. For reviewers who provided code snippets in code reviews, most of them (i.e., 75.0\% in OpenStack and 83.2\% in Qt) provided snippets to \textit{Improve Code Implementation}, which is also the most common purpose regarding the number of code snippet comments. In OpenStack, the percentages of reviewers providing code snippet suggestions for the other three purposes (i.e., \textit{Following Code Style}, \textit{Correcting Code}, and \textit{Complementing Code Implementation}) are very close, and reviewers who provided code snippet comments for the purpose of \textit{Providing Context} account the least (22.4\%). In Qt, the proportion of reviewers who provided code snippet comments aiming at \textit{Following Code Style} account the least (25.7\%), while the percentages of reviewers who provided code snippet comments for the remaining four purposes (i.e., \textit{Correcting Code}, \textit{Complementing Code Implementation}, \textit{Elaborating}, and \textit{Providing Context}) are comparable.

\begin{table}[htbp]
\centering
\renewcommand\arraystretch{1.5}
\caption{Counts and percentages of reviewers who had provided code snippets in code reviews for different purposes}
\label{Counts and percentages of reviewers who had provided code snippets in code reviews with different purposes}
\begin{tabular}{lm{2.5cm}<{\centering}m{2.5cm}<{\centering}}
\hline
\textbf{Detailed Purpose}          & \textbf{\#Reviewers in OpenStack (\%)}  & \textbf{\#Reviewers in Qt (\%)} \\ \hline
Improving Code Implementation      & 57 (75.0\%)                            & 84 (83.2\%) \\
Following Code Style               & 24 (31.6\%)                            & 26 (25.7\%) \\
Correcting Code                    & 26 (34.2\%)                            & 43 (42.6\%) \\
Complementing Code Implementation  & 25 (32.9\%)                            & 39 (38.6\%) \\ \hline
Elaborating                        & 21 (27.6\%)                            & 47 (46.5\%) \\
Providing Context                  & 17 (22.4\%)                            & 45 (44.6\%) \\ \hline
\end{tabular}
\end{table}

\subsubsection{Feedback on the Findings of RQ2}
According to the results of how often developers provide code snippets in code reviews of the survey study, 4 participants (6.3\%) answered that they never provided code snippets in review comments. So we analyzed the responses of the remaining 59 participants to investigate the purposes of industrial developers providing code snippets. The survey question (SQ3) in the questionnaire provides the six detailed purposes of code snippets obtained from the repository mining study (see Section~\ref{subsubsec: results_of_RQ2}) as candidate options for participants to choose. Note that the participants can select multiple purposes when answering this survey question, and they can also fill in the ``Other'' field to express other purposes of providing code snippets that are not covered by the candidate options. In the 59 responses, 3 participants completed the ``Other'' field. However, we decided to exclude their answers to the ``Other'' field as these 3 participants did not explain their other purposes clearly. In the ``Other'' field, they just expressed their opinions about using code snippets in code reviews and did not highlight the point, as one of them stated ``\textit{nearly only if writing code is more readable and more explanatory than normal text}'', which is about the characteristics of code snippets instead of a clear purpose of providing them.

\begin{figure}[htbp]
	\centering
	\includegraphics[width=1.0\linewidth]{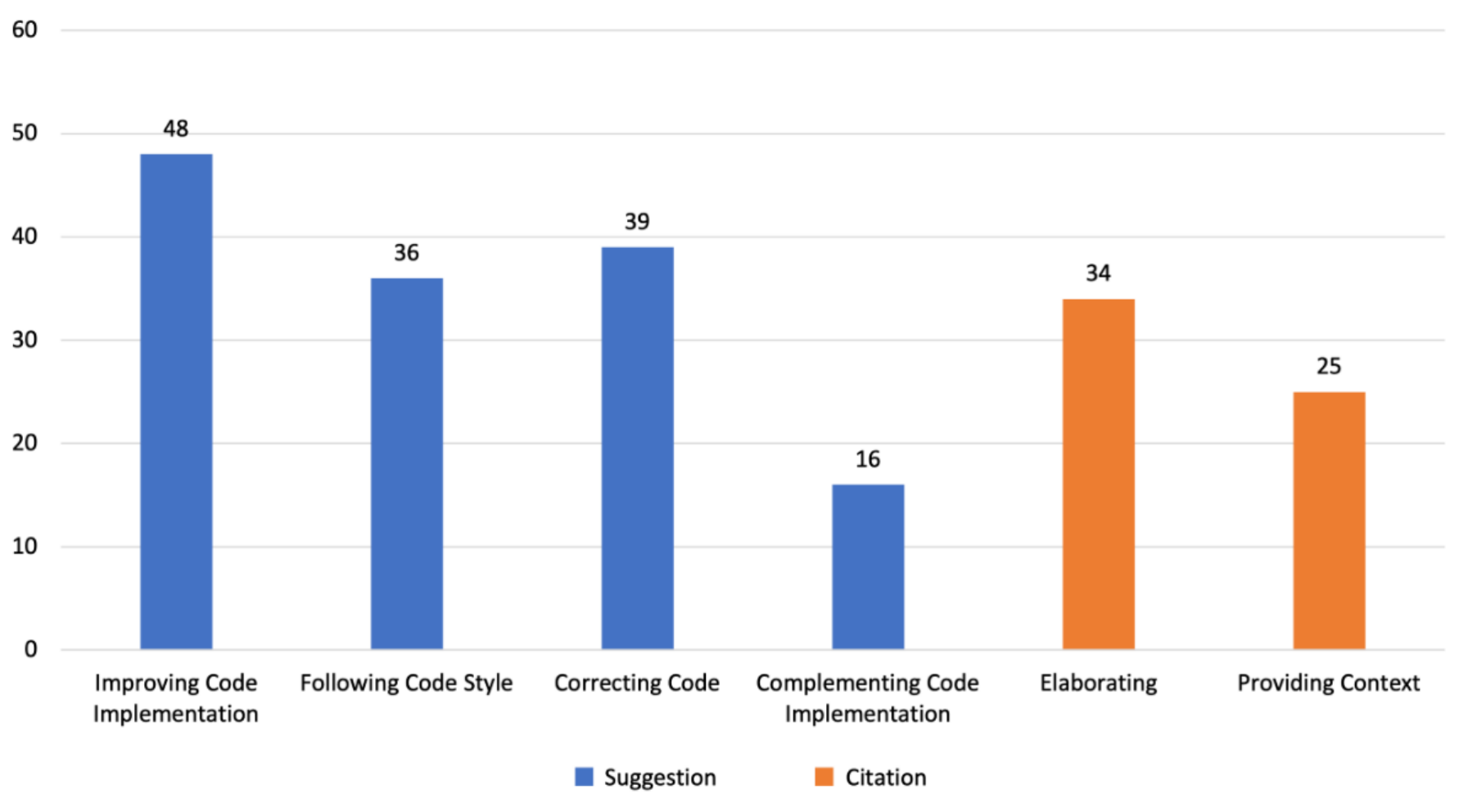}
	\caption{Counts of detailed purposes of using code snippets in code reviews from the industrial survey}
	\label{fig: Feedback_on_Findings_of_RQ2}
\end{figure}

Fig.~\ref{fig: Feedback_on_Findings_of_RQ2} presents the feedback on our findings of RQ2. It shows that the most common purpose of participants providing code snippets in code reviews is \textit{Suggestion}. In total, 45 participants said that they would provide code snippets for \textit{Improving Code Implementation}, accounting for 76.2\%, making \textit{Improving Code Implementation} the most common detailed purpose for industrial developers to provide code snippets in code reviews. This result conforms to the repository mining study results (i.e., the proportion of reviewers who had provided code snippets for different purposes). The counts and percentages of \textit{Correcting Code}, \textit{Following Code Style}, and \textit{Elaborating} are very close, accounting for 66.1\% (39), 61.0\% (36), and 57.6\% (34) respectively. Moreover, 25 participants selected \textit{Providing Context} as their purpose of using code snippets in code reviews, accounting for 42.4\%. Only 16 participants said that they would provide code snippets with the purpose of \textit{Complementing Code Implementation} when conducting code reviews as reviewers, accounting for 27.1\% merely.

\noindent\begin{center}
		\begin{tcolorbox}[colback=black!5, colframe=black!20, width=1.0\linewidth, arc=1mm, auto outer arc, boxrule=1.5pt]                       
            {{\textbf{RQ2 Summary:} We identified the purposes of reviewers providing code snippets in code reviews and categorized them into two high-level categories: \textit{Suggestion} and \textit{Citation}. We further identified six types of detailed purposes under the two high-level categories, including \textit{Improving Code Implementation}, \textit{Following Code Sytle}, \textit{Correcting Code}, \textit{Complementing Code Implementation}, \textit{Elaborating}, and \textit{Providing Context}. Furthermore, we refined the \textit{Improving Code Implementation} purpose by categorizing six specific purposes under this category.} The results show that the main purpose of using code snippets by reviewers in code reviews is \textit{Suggestion}, which accounts for 83.9\% in total. Besides, most code snippets are provided for the purpose of \textit{Improving Code Implementation} to enhance code maintainability.}
		\end{tcolorbox}
\end{center}

\subsection{RQ3: The Developers’ Acceptance of Code Snippet Suggestions}
\label{subsec:results_of_RQ3}

To answer RQ3, we explored how developers treat code review comments containing code snippets for the purpose of \textit{Suggestion}, and we found that developers either accept, ignore, or do not accept code snippet suggestions. Then we investigated (1) the distribution of developers accepting, ignoring, or not accepting code snippet suggestions and (2) the acceptance rate of code snippet suggestions for the four detailed purposes of \textit{Suggestion} presented in the results of RQ2 (see Section~\ref{subsec:results_of_RQ2}).

\vspace{10pt}

\noindent\textbf{The distribution of developers accepting, ignoring, or not accepting code snippet suggestions}

\begin{figure}[htbp]
	\centering
	\includegraphics[width=1.0\linewidth]{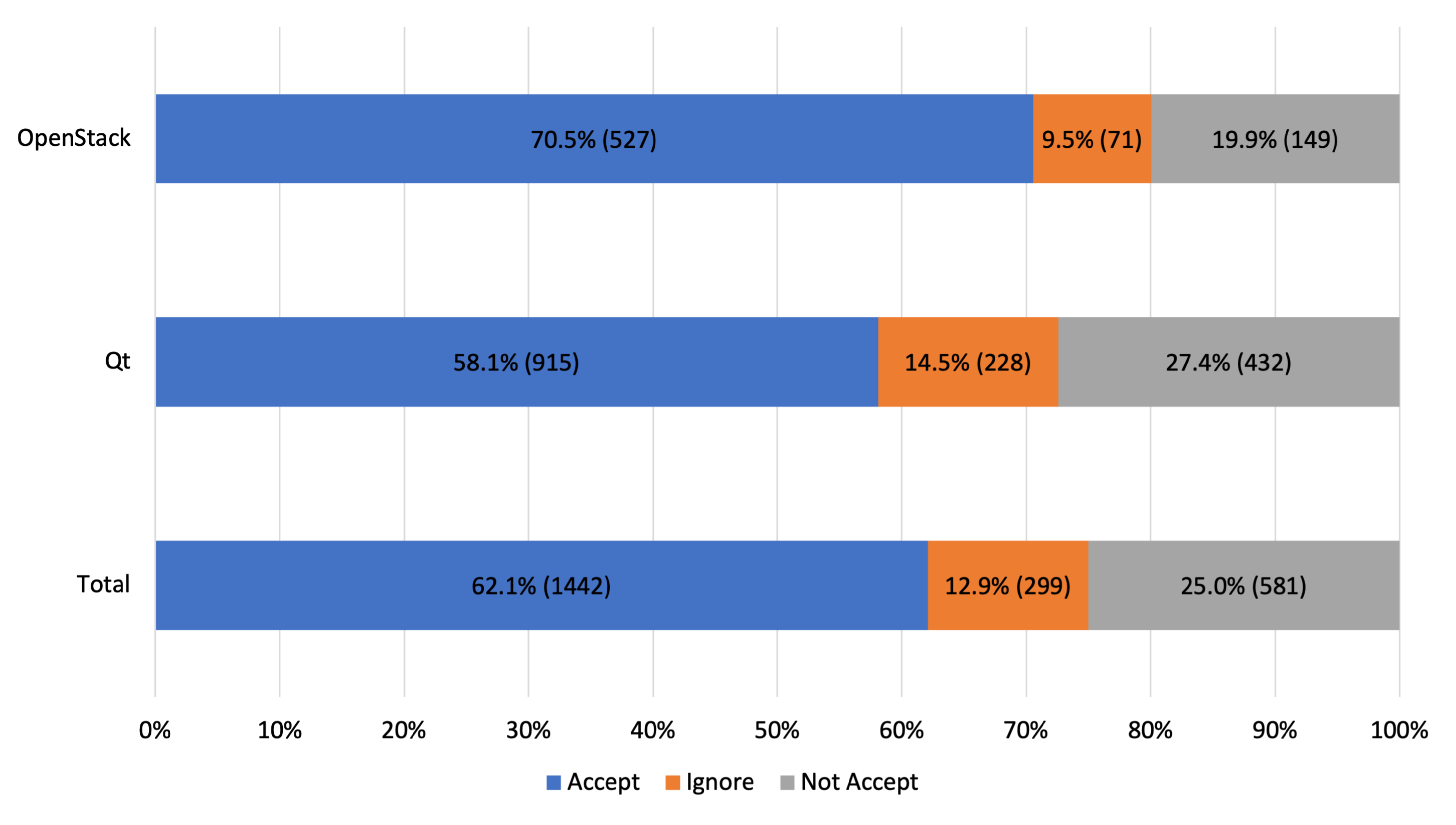}
	\caption{Distribution of how developers treat code snippet suggestions in code reviews}
	\label{fig: Distribution of How Developers Treat Code Snippet Suggestions in Code Reviews}
\end{figure}

The distribution of how developers treat code snippet suggestions in code reviews is shown in Fig. \ref{fig: Distribution of How Developers Treat Code Snippet Suggestions in Code Reviews}. In OpenStack and Qt, we got 2,322 code snippet suggestions, 1,442 of which were accepted by developers, accounting for 62.1\%. However, 12.9\% of the code snippet suggestions were just ignored by developers, who neither responded to these review comments nor made corresponding code changes. 581 code snippet suggestions were not accepted by developers, accounting for 25.0\%. In the OpenStack and Qt communities, the acceptance of code snippet suggestions is 70.5\% and 58.1\%, respectively. As a whole, most code snippet suggestions (more than 60\%) were accepted by developers, and about 40\% of code snippet suggestions were just ignored or not accepted. This finding suggests that code snippets can serve as an effective way when reviewers provide suggestions for developers in code reviews.

\begin{figure}[htbp]
	\centering
	\includegraphics[width=1.0\linewidth]{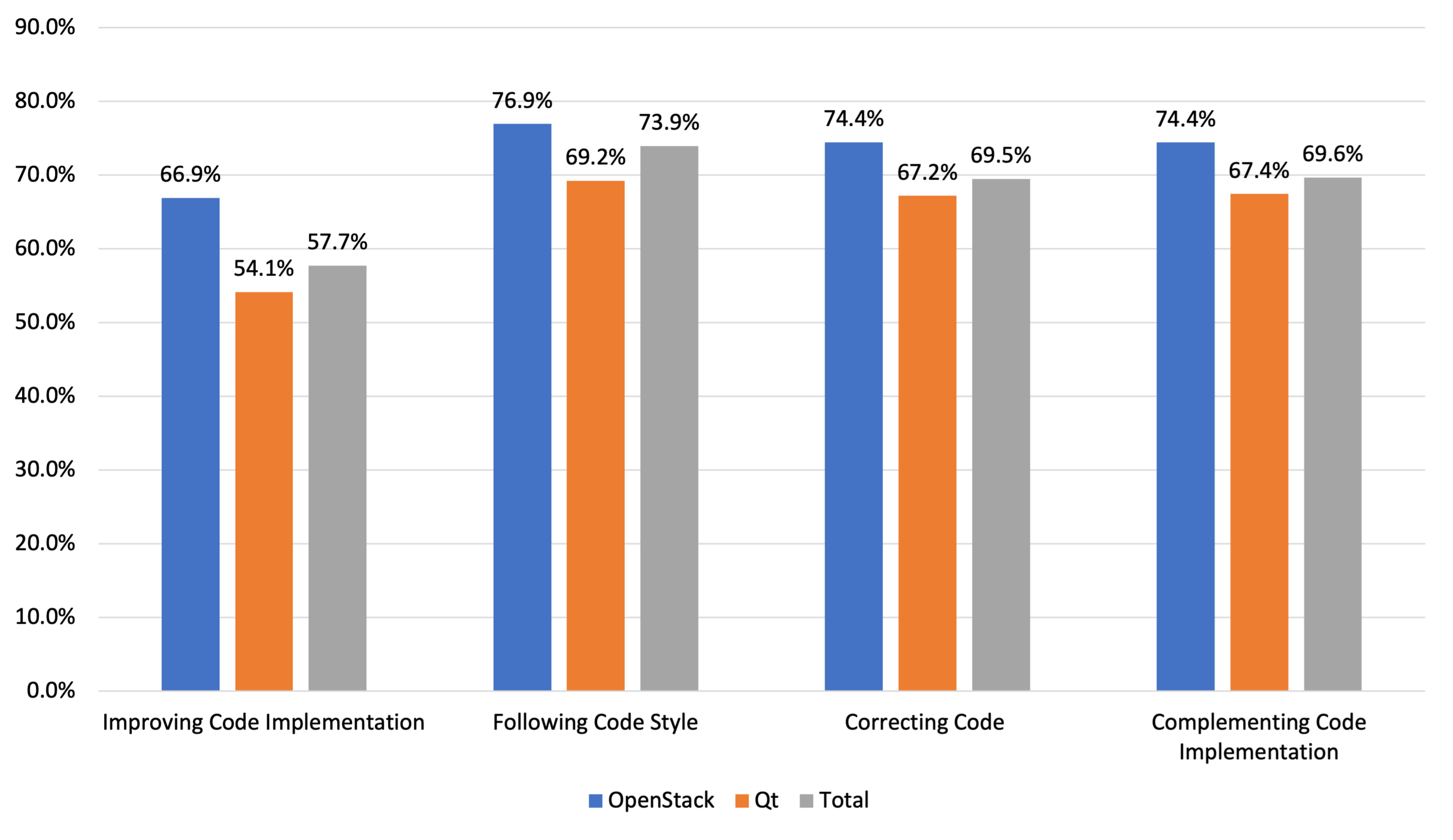}
	\caption{Acceptance rate of code snippets in code reviews for four detailed purposes of Suggestion}
	\label{fig: Acceptance rate of code snippets in code reviews for four detailed purposes of Suggestion}
\end{figure}

\vspace{10pt}

\noindent\textbf{The acceptance rate of code snippet suggestions for four detailed purposes of \textit{Suggestion}}

We further investigated the acceptance rate of four detailed \textit{Suggestion} purposes as presented in Fig. \ref{fig: Acceptance rate of code snippets in code reviews for four detailed purposes of Suggestion}. Overall, the result shows that the acceptance rate of \textit{Following Code Style} (73.9\%) is the highest, while the acceptance rate of \textit{Improving Code Implementation} (57.7\%) is the lowest. Besides, the acceptance rates of \textit{Correcting Code} and \textit{Complementing Code Implementation} are very close, with a minor difference 0.1\% between them. 

\noindent\begin{center}
		\begin{tcolorbox}[colback=black!5, colframe=black!20, width=1.0\linewidth, arc=1mm, auto outer arc, boxrule=1.5pt]                       
            {{\textbf{RQ3 Summary:} 
            In general, most code snippet suggestions (62.1\%) were accepted by developers}. Among the four detailed \textit{Suggestion} purposes, the acceptance rate of \textit{Following Code Style} is the highest, while the acceptance rate of \textit{Improving Code Implementation} is the lowest.}
		\end{tcolorbox}
\end{center}

\subsection{RQ4: The Reasons that Developers do not Accept Code Snippet Suggestions}
\label{subsec:results_of_RQ4}
According to the results of RQ3, 581 code snippet suggestions made by reviewers were not accepted by developers. To answer RQ4, we further analyzed these not accepted code snippet suggestions through the discussions between developers and reviewers. Based on the contextual information around the review comments, we identified the reasons why developers do not accept code snippet suggestions as listed in Table \ref{The reasons that developers do not accept code snippet suggestions in code reviews}.
\begin{table}[htbp]
\renewcommand\arraystretch{1.5}
\centering
\caption{Counts and percentages of the reasons that developers do not accept code snippet suggestions in code reviews}
\label{The reasons that developers do not accept code snippet suggestions in code reviews}
\begin{tabular}{lcc}
\hline
\textbf{Reason}                                               & \textbf{\#}       & \textbf{\%}  \\ \hline
Difference in the opinions between developers and reviewers   & 136               & 31.1\%       \\ 
Reviewer's suggestion is flawed                               & 105               & 24.0\%       \\ 
Adopt other proposed suggestion                               & 70                & 16.0\%       \\
Consider as an optional suggestion                            & 41                & 9.4\%        \\ 
Consider as a future plan                                     & 35                & 8.0\%        \\ 
Need to be discussed                                          & 15                & 3.4\%        \\ 
Avoid extra code complexity                                   & 10                & 2.3\%        \\ 
Improve code readability                                      & 9                 & 2.1\%        \\ 
Developer removed the relevant code                           & 9                 & 2.1\%        \\ 
Keep code consistency                                         & 5                 & 1.1\%        \\ 
Conflict with other patch                                     & 2                 & 0.5\%         \\ \hline       
\end{tabular} 
\end{table}

Many developers did not clarify the reasons behind their non-acceptance of code snippet suggestions. Out of the 581 unaccepted code snippet suggestions, 144 were not accepted due to unknown reasons. Of the remaining 437 unaccepted code snippet suggestions, \textit{difference in the opinions between developers and reviewers} (31.1\%) and \textit{reviewer's suggestion is flawed} (24.0\%) are the two main causes for the non-acceptance of code snippet suggestions. Since the differences in coding preferences, coding ability, software development experience, and understanding of current code changes, developers may hold different opinions against reviewers' code snippet suggestions. For example, in the following review comment, the developer thought that changing the code according to the reviewer's suggestion was unworthy, and he decided to keep the current code and rejected the reviewer's suggestion.

\begin{qoutebox}{white}{}
    \textbf{Link:} \url{http://alturl.com/hb2m6} \\
    \textbf{Reviewer:} ``Suggestion: add a method to remove the version to the Import (moti- vation: shorter, this fix might be needed elsewhere). So then this would be shortened to: [code snippet]''\\
    \textbf{Developer:} ``Import currently has no non-const methods so let’s keep it that way... Could use reference here, too, but the minimal perf gain from that in literally meaningless here, so not worth changing.''
\end{qoutebox}

Another main reason for developers not accepting reviewers' code snippet suggestions is \textit{reviewer's suggestion is flawed}. Sometimes, reviewers miss the context and provide unsuitable solutions to current code change (see the example below). Sometimes, reviewers proposed code snippets that have syntax or logical errors. Therefore, developers pointed out the mistakes in the code snippet suggestions, and rejected the suggestions.


\begin{qoutebox}{white}{}
    \textbf{Link:} \url{http://alturl.com/6kb4v} \\
    \textbf{Reviewer:} ``Could we use this on L401? [code snippet]''\\
    \textbf{Developer:} ``Why? image\_chunks is intentionally initialized here so it does not get filled in L381-392 (via the glance API call)''\\
    \textbf{Reviewer:} ``Yeah apologies I somehow missed that.''
\end{qoutebox}

Some code snippet suggestions were rejected because developers \textit{adopted other proposed solutions}. The adopted solutions may come from other reviewers, or from the developers themselves. Some code snippet suggestions were \textit{considered as an optional suggestion} which are not necessary to follow or \textit{considered as a future plan} which were not accepted now but might be accepted in the follow-up patchsets. Another reason why developers do not accept code snippet suggestions is because they thought that whether to modify the current code based on the code snippets provided by reviewers \textit{needs to be discussed}. For the reasons to \textit{avoid extra code complexity} and \textit{improve code readability}, developers do not accept reviewers' code snippet suggestions to make code simpler and more readable (see the two examples below):

\begin{qoutebox}{white}{}
    \textbf{Link:} \url{http://alturl.com/qnnou} \\
    \textbf{Reviewer:} ``type hints? Note that you can avoid circular imports with something like e.g. 'nova.network.neutron.API' by doing the following: [code snippet]''\\
    \textbf{Developer:} ``Note that I don’t want to be reluctant to adding type hints here, but I think we are adding extra code complexity for an unnecessary need...''
\end{qoutebox}

\begin{qoutebox}{white}{}
    \textbf{Link:} \url{http://alturl.com/cbrvr} \\
    \textbf{Reviewer:} ``NIT (just a possibly stupid suggestion): [code snippet]''\\
    \textbf{Developer:} ``I will keep as-is now for readability. These methods have a long name so doing 1 liners with them makes the indentation fun.''
\end{qoutebox}

In some review comments, \textit{developers removed the relevant code} so they did not accept reviewers' code snippet suggestions:

\begin{qoutebox}{white}{}
    \textbf{Link:} \url{http://alturl.com/gu5cn} \\
    \textbf{Reviewer:} ``by the way whil i like the operator module if this is all you are using it for i would just do [code snippet]''\\
    \textbf{Developer:} ``Again, just code motion, this is actually dropped in the follow up series by [URL].''
\end{qoutebox}

\textit{Keeping code consistency} and \textit{conflicting with other patch} are the least mentioned reasons by developers, accounting for 1.1\% and 0.5\% respectively.

\noindent\begin{center}
		\begin{tcolorbox}[colback=black!5, colframe=black!20, width=1.0\linewidth, arc=1mm, auto outer arc, boxrule=1.5pt]                       
            {{\textbf{RQ4 Summary:} We identified 11 categories of reasons why developers do not accept reviewers' code snippet suggestions, among which \textit{difference in the opinions between developers and reviewers} is the major reason.}} 
		\end{tcolorbox}
\end{center}

\subsection{RQ5: The Scenarios in which Reviewers Provide Code Snippets}
\label{subsec:results_of_RQ5}
A total of 36 participants (57.1\%) answered the open-ended question (SQ4) to provide the scenarios of using code snippets in code reviews. Table \ref{Feedback on scenarios in which reviewers often provided code snippets in code reviews} presents the 11 scenarios in which reviewers provide code snippets when conducting code review collected from the feedback of industrial developers.
\begin{table}[htbp]
\renewcommand\arraystretch{1.5}
\centering
\caption{Counts of the scenarios in which reviewers provide code snippets in code reviews}
\label{Feedback on scenarios in which reviewers often provided code snippets in code reviews}
\begin{tabular}{lc}
\hline
\textbf{Scenario}                                     & \textbf{\#} \\ \hline
When code is more illustrative than words             & 27         \\ 
Current code is suboptimal                            & 9          \\ 
Advise developers not to deviate from code style      & 5          \\ 
Provide standard code snippet examples                & 5          \\
Show algorithms, standard libraries, or tools         & 3          \\ 
Guide new contributors                                & 2          \\
Avoid potential communication issues                  & 2          \\ 
Help developers know the target code elements         & 2          \\ 
Elaborate the impact of code changes on user behavior & 1          \\ 
Ask about the relationship between code snippets      & 1          \\ 
Refer to another code example                         & 1          \\ \hline       
\end{tabular}
\end{table}

Most participants expressed that they would like to provide code snippets in review comments \textit{when code is more illustrative than words} (27/36, 75.0\%). For example, when a large block of code has logical errors, it is easier to describe the errors by using code snippets rather than words. As one participant stated ``\textit{I think code snippets can be clearer than describing the fix/improvement suggestion in some situations}''. 

Nine of the participants mentioned that when \textit{current code is suboptimal} (e.g., when current code ``\textit{is unnecessarily complex or hard to follow}''), they would provide code snippets to help developers improve code quality. Five participants said that they would using code snippets to \textit{advise developers not to deviate from code style}. 
Besides, five other participants mentioned that they would \textit{provide standard code snippet examples} to ``\textit{help developers better understand the suggestions in code review comments}''.

Three participants said that they would use code snippets to \textit{show algorithms, standard libraries, or tools} for developers. One of the three participants described the process in detail, ``\textit{I wonder if the programmer has implemented the function himself instead of using the standard library. In these cases, only writing is not enough, and for better understanding, a sample code should also be written in the review. With an example, it can be said that the standard library can be used in this way and the programmer’s time can be saved in this way}''. Moreover, two participants indicated that ``\textit{when a difficult issue arises}'' and the new contributors ``\textit{are not entirely able to identify why the tests are not passing}'', they would provide code snippets in review comments to \textit{guide new contributors}. Two other participants said that they would consider using code snippets in code reviews to \textit{avoid potential communication issues}, which can accelerate the process of problem solving.

Some participants indicated that they would provide code snippets in code reviews when there is a need to \textit{help developers know the target code elements}, \textit{elaborate the impact of code changes on user behavior}, \textit{ask about the relationship between code snippets}, and \textit{refer to another code example}. These scenarios are similar to the cases of providing code snippets with the purpose of \textit{Elaborating} or \textit{Providing Context} in the results of RQ2 (see Section~\ref{subsubsec: results_of_RQ2}).


\noindent\begin{center}
		\begin{tcolorbox}[colback=black!5, colframe=black!20, width=1.0\linewidth, arc=1mm, auto outer arc, boxrule=1.5pt]                       
            {{\textbf{RQ5 Summary:} Among the 63 responses from industrial practitioners, 36 participants gave the scenarios in which reviewers provide code snippets in code reviews. Based on the responses, we identified 11 categories of scenarios, among which over 80\% participants mentioned that they would use code snippets in review comments \textit{when code is more illustrate than words}.
            }}
		\end{tcolorbox}
\end{center}

\subsection{RQ6: The Developers' Attitudes towards Code Snippets}
\label{subsec:results_of_RQ6}
\begin{figure}[htbp]
	\centering
	\includegraphics[width=1.0\linewidth]{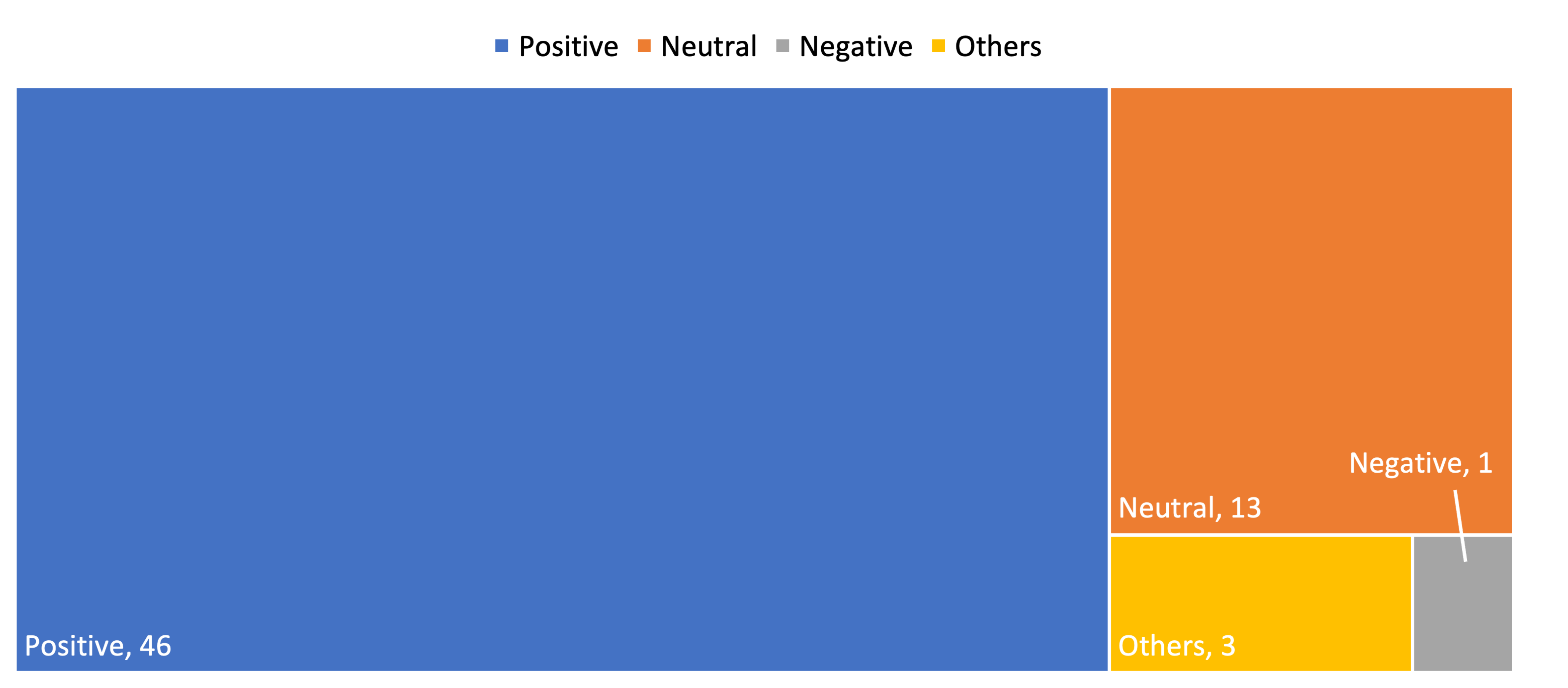}
	\caption{Distribution of the attitudes of industrial practitioners towards provided code snippets in code reviews}
	\label{fig: Industrial developers' attitudes towards provided code snippets in code reviews}
\end{figure}

The attitudes of industrial practitioners towards provided code snippets in code reviews is shown in Fig.~\ref{fig: Industrial developers' attitudes towards provided code snippets in code reviews}. Out of the 63 responses, 46 participants (73.0\%) expressed positive attitudes, thinking that code snippet comments are a good thing. 
Besides, 13 participants held a neutral attitude towards provided code snippets, and they did not care whether review comments contain code snippets. 

Only one participant mentioned that using code snippets ``\textit{is not beneficial for someone else to solve an individual’s problems for him or her}'', and is negative about provided code snippets in code reviews.

Three participants filled in the ``Other'' field to answer the survey question. One of them commented, ``\textit{It depends, if it makes the review process faster, good. If it’s somebody telling you ‘you should be doing something else’ then bad}''. Another participant stated that, ``\textit{Generally positive, as long as it facilitates communication}''. The attitudes of these two participants towards provided code snippets in review comments depend on whether the code snippets are good for code review process. If providing code snippets can facilitate code review, they would hold a positive attitude. Otherwise, they would be negative. One other participant mentioned that ``\textit{I’d normally consider it to be a problem if there were things in the review that could be solved by code snippets}''. At the same time, this participant wrote that the members of their development team were experienced, and they had worked in software industry for 6$\sim$40 years, so there were few needs for code snippets in code reviews.

\noindent\begin{center}
		\begin{tcolorbox}[colback=black!5, colframe=black!20, width=1.0\linewidth, arc=1mm, auto outer arc, boxrule=1.5pt]                       
            {{\textbf{RQ6 Summary:} For provided code snippets by reviewers in code reviews, most developers (73.0\%) hold a positive attitude. Besides, some developers' attitudes towards provided code snippets in review comments depend on whether these code snippets are beneficial to the review process. If the code snippets are conducive, the developers will be positive about them.}}
		\end{tcolorbox}
\end{center}

\subsection{RQ7: The Characteristics of Code Snippets Developers Expect Reviewers to Provide}
\label{subsec:results_of_RQ7}
Out of the 63 participants, 38 participants (60.3\%) provided valid answers to the open-ended question (SQ6). Table \ref{Feedback on characteristics of code snippets that developers expect reviewers to provide in code review} shows the 4 categories of characteristics of code snippets that developers expect reviewers to provide in review comments.

\begin{table}[htbp]
\renewcommand\arraystretch{1.5}
\centering
\caption{Counts of the characteristics of code snippets that developers expect reviewers to provide in code reviews}
\label{Feedback on characteristics of code snippets that developers expect reviewers to provide in code review}
\begin{tabular}{m{3.5cm}m{6.8cm}c}
\hline
\textbf{Characteristic}          & \textbf{Example}                                                                        & \textbf{\#} \\ \hline
Understandable                   & \textit{Pseudocode to help others understand my idea}                                   & 39          \\ 
Fitting into existing code       & \textit{Generally simple examples which can be fit into the existing code with ease}    & 28          \\ 
Providing a better solution      & \textit{Sample code that can provide a better solution}                                 & 3          \\ 
Scenario-specific                & \textit{Scenario-specific code snippets}                                                & 1          \\ \hline       
\end{tabular} 
\end{table}

From Table \ref{Feedback on characteristics of code snippets that developers expect reviewers to provide in code review}, we can find that \textit{undetstandable} and \textit{fitting into existing code} are the major characteristics developers want most in provided code snippets. More specifically, most developers expect \textit{understandable} code snippets, and they hope that the code snippets should be simple, easy to read, provide detaied context, and highlight the point. As one participant stated, ``\textit{In my opinion, the desired code should be as simple as possible. In this way, the programmer can quickly understand what to do, away from the sidelines}''. 28 participants indicated that provided code snippets should be \textit{fitting into existing code}. They hope reviewers to provide functional code snippets which can execute easilty. As one participant mentioned, ``\textit{Preferably be possible to copy \& paste directly as a replacement to the commented code''.}

Few participants also stated other characteristics (i.e., \textit{providing a better solution} and \textit{scenario-specific}) that they expected reviewers to provide with code snippets. We also found that some participants did not care about the characteristics of code snippets in code reviews. They just wanted reviewers to provide code snippets that could promote the code review process, as one participant wrote ``\textit{I have no preferences for this: I trust reviewers to use code snippets where this is the most effective way to communicate their review comment}''.

\noindent\begin{center}
		\begin{tcolorbox}[colback=black!5, colframe=black!20, width=1.0\linewidth, arc=1mm, auto outer arc, boxrule=1.5pt]                       
            {{\textbf{RQ7 Summary:} We identified 4 types of characteristics of code snippets that developers expect reviewers to provide in code reviews, and we found that most developers expect provided code snippets in code reviews to be \textit{understandable} and \textit{fitting into existing code}}.}
		\end{tcolorbox}
\end{center}

\section{Discussion}
\label{sec:discussion}
In this section, we present a broader discussion of our empirical observations based on the results of both the exploratory and survey study.

\textbf{Developers should follow the best code conventions when programming and reviewers should pay more attention to code style issues during code review:} According to the results of RQ2, \textit{Following Code Style} is one of the main purposes that reviewers provide code snippets in review comments. This finding indicates that developers may be unfamiliar with the code style used in the projects and thus may lead into code style issues. \cite{miltiadis2014learning} found that one third of change reviews include feedback regarding code conventions, and developers are often unaware of the conventions. We suggest developers to write code in adherence to best code conventions to standardize the code formatting and reduce code style issues, which can improve consistency of code and collaboration within development teams. According to the results of RQ3, the acceptance rate of \textit{Following Code Style} is the highest among the four detailed \textit{Suggestion} purposes, which implies that developers and reviewers can easily reach a consensus towards suggestions related to code style. However, according to the proportions of reviewers who had provided code snippet comments for different purposes (Table~\ref{Counts and percentages of reviewers who had provided code snippets in code reviews with different purposes} in Section \ref{sub_results_of_rq2}), the ratio of reviewers providing suggestions aiming at \textit{Following Code Style} ranks the least, which indicates that only a small part of reviewers notice the inconsistency between the style of reviewed code and the best code conventions. We suggest reviewers to pay more attention to code style issues during code review, and this will improve the quality of code, just as one participant mentioned, ``\textit{It is very important that all team members use the same style code. I usually advise everyone not to deviate from this style code}''. Besides, automation tools that examine whether the source code follows the project's code conventions can also be highly beneficial for software development. These measures can significantly reduce reviewers' workload on commenting code style issues and accelerate code review process.

\textbf{OSS development teams can use code snippets in review comments to guide new contributors:} Popular open source projects receive review contributions from a diverse group of developers, including many who have limited or no previous engagement with the project \citep{vincent2015will}. \cite{jennifer2013impression} found that OSS developers rely on information about detailed traces of an individual’s project-related activities to inform their decisions on how to interact with new, unknown contributors to their projects. Code snippets can also be used to convey necessary information to novel contributors during code review process. According to the results of RQ5, two participants expressed their willingness to provide code snippets in review comments in order to \textit{guide new contributors}. One of them said, ``\textit{Most often, I provide snippets to provide context to newer contributors on items such as code style checks which are resulting in their tests failing, or when a difficult issue arises and they are not entirely able to identify why the tests are not passing. Sometimes this is to more elaborate or make the code more defensive, but generally more experienced developers tend to do this inherently so the need is less}''. The use of code snippets is a valuable tool to guide and inform new contributors during code review. Experienced developers require fewer code snippet comments in code reviews compared to new contributors as experienced developers tend to write high-quality code naturally. Nevertheless, we recommend that knowledgeable developers within OSS development teams can utilize code snippets in code reviews to facilitate the onboarding and contribution process for new team members, fostering collaboration and growth of these projects.

\textbf{Reviewers can more frequently provide code snippet suggestions in code reviews to encourage developers' acceptance:} Overall, 62.1\% developers accepted reviewers' code snippet suggestions according to the results of RQ3, revealing that most developers changed existing code to code snippets provided by reviewers in the following patches. Such a high level of acceptance rate of reviewers' code snippet suggestions demonstrates that code snippets can greatly assist developers in understanding reviewers' suggested code changes as code snippets show concrete and actionable solutions. Therefore, developers can make necessary adjustments to current code accordingly. These code snippet suggestions are valuable for enhancing code quality attributes, promoting consistency in code style, ensuring code correctness, and preserving code integrity (see the purposes of providing code snippet suggestions in Section \ref{sub_results_of_rq2}). By more frequently providing code snippet suggestions during code review process, reviewers not only enhance the efficiency of communications, but also deepen developers' understanding of how to make code changes, ultimately resulting in developers' higher acceptance rate of reviewers' code snippet suggestions.

\textbf{Code snippets can serve as an effective way for communication during code review:} \cite{gonalves2023competencies} have focused on the competencies developers need to execute code review, and their study results call for more research on how to support and develop reviewers' potential to communicate effectively during code review. Our study results show that code snippets can be used to improve the effectiveness of communication between reviewers and developers. According to the responses from our survey study, most participants thought that in some scenarios code snippets provide a quicker, easier, and unambiguous way to express their opinions and suggestions. As one participant mentioned that, ``\textit{Many times, presenting a piece of code works better than explaining it for hours, and the programmer quickly understands what to do}''. Some other participants believed that using code snippets in code reviews could help developers evaluate the suggestions made by reviewers, as one response wrote that ``\textit{It also makes it very easy to evaluate the suggestion by the developer}''. According to the results of RQ6, some participants even mentioned that they would provide code snippets in code reviews to \textit{avoid potential communication issues}, indicating that they thought that providing code snippets in review comments could help them convey needed information for code review more accurately.

\textbf{Purposeful code snippets in code reviews should be understandable and easy to fit into existing code:} \cite{bacchelli2013expectations} explored code review comments in practice, and they found that the key of any review is code and change understanding. Our study results corroborate this to some extent, as the results of RQ7 highlight the importance of the understandability of provided code snippets in review comments. Code snippets should be written by reviewers in a way that makes them easy for developers to understand. They should be clear and concise, without unnecessary complexity or ambiguity, making it easy for developers to grasp its significance within the context of the code review. Otherwise, provided code snippets that are not relevant may unnecessarily divert developers' attention and add extra workload. As one participants said, ``\textit{Code snippets should be short and to the point; not containing boilerplate which would be needed to run the code in isolation; as that distracts from the main point raised in the review}''. Meanwhile, provided code snippets should be easy to \textit{fit into existing code}. In other words, the code snippets should be ``\textit{syntactically correct and executable}'', compatible with the code style and functionality of the codebase, ensuring that incorporating the code snippets into the existing code will not introduce errors or inconsistencies. RQ4 investigates the reasons why developers do not accept code snippet suggestions, and \textit{reviewers' suggestion is flawed} accounts for nearly 25\%, ranking it second, which indirectly confirms that provided code snippets should be correct.


\textbf{Developers' purposes of providing code snippets in code reviews need further investigation:} According to the results of RQ1, in addition to reviewers, developers provide code snippets in review comments as well. The results of RQ2 provide reviewers' purposes of providing code snippets in code reviews. However, we did not explore the purposes of developers providing code snippets in code reviews, which could be different from the purposes of reviewers and can be further investigated and compared in the future. In code review process, developers also need to provide information to meet reviewers' needs. If this information is readily available, reviewers can focus on verifying and improving the code they are reviewing, rather than spending time and effort on inquiring and collecting missing information \citep{pascarella2018information}. Besides, the results of RQ3 present the acceptance rate of code snippet suggestions, showing that most suggestions were accepted by developers (61.9\%). This finding provides an empirical basis for researchers to further explore how the inclusion of code snippets in reviewers' \textit{suggestion} comments has an impact on developers' reactions.

\section{Threats to Validity}
\label{sec:threats}

Given the empirical nature of our study, we discuss several threats to the validity of this work according to the guidelines proposed by \cite{runeson2009guidelines}, and how these threats were partially mitigated in our study.


\textbf{Construct Validity:} In this work, we depended on human activities, including data labelling and data extraction \& analysis, which would introduce personal bias. To reduce this threat, each step in the aforementioned human activities was conducted by two authors and a third author was involved to discuss and resolve the conflict in case of disagreement. Moreover, we also conducted a pilot data labelling to make sure that the two researchers achieved a consensus on what are code snippets in this study, which could also partially alleviate this threat. In RQ2, we investigated the purposes of reviewers providing code snippets in code reviews. 10 review comments contain multiple code snippets, from which we only extracted the most significant purpose of these code snippets. This poses some threat to the construct validity of the results of RQ2. However, given the small number of review comments (0.31\%, 10/3213) that contain multiple code snippets, we believe that this threat to the construct validity is minimal.

Another threat to the construct validity of this study is that we used mostly closed-ended questions in the industrial survey, which may affect the richness of the responses collected from the participants. However, as argued by \cite{reja2003openended}, open-ended questions have several disadvantages compared with closed-ended questions. For example, much long time to fill out the questionnaire might make participants do not participate in the survey at all. Participants may provide poor answers or even just skip when answering open-ended questions. Due to the above disadvantages, we chose to mainly use closed-ended questions in our survey. For some of the closed-ended questions, we also provided the ``Other'' field so that participants can fill in their own opinions if existing options do not cover their thoughts. Furthermore, to help participants better understand the open-ended questions in the survey, we provided two examples for each question. During the data analysis, we found that some participants only agreed with the provided examples without providing additional answers, which indicates that the provided examples may restrict participants from providing their own answers to the open-ended questions, thus affecting the richness of answers. Besides, another threat is that some of the responses from participants are written in Chinese, and translating the raw data from Chinese to English may lead to information lost or corruption. The two authors who extracted and analyzed the Chinese responses are native Chinese speakers, and the third author who is a native Chinese speaker as well was asked to check and refine the translation, which partially minimizes this threat.

The last threat is concerning the size of our dataset. We collected 63 responses from our industrial survey, and we acknowledge that the small number of responses may threaten the validity of our findings. Therefore, we conjecture that we could obtain more convincing results by inviting more developers with code review experiences from more diverse communities to participate in the survey, which is also our next step.


\textbf{External Validity:} We selected the four most active projects from the OpenStack and Qt communities since these two communities have made a serious investment in code review for many years and have been widely used in many studies related to code review. We argue that the selected communities and projects are representative and can increase the generalizability of our study results. 

In terms of the industrial survey, we invited developers from the OpenStack and Qt communities collected from our dataset, developers from well-known software companies in China, and developers from professional software development groups, which partially increases the external validity of the survey results. But we admitted that the findings of this study may not be generalized to all developers. In the future, we plan to invite more developers from various development groups (e.g., inner source development) to expand the scope of the industrial survey.

\textbf{Reliability:} To improve the reliability, we made a research protocol with detailed procedure, which was discussed and confirmed by all the authors. Besides, all of the empirical steps in our study, including the data mining process, data labelling, and data extraction and analysis, were conducted and discussed by three authors. Furthermore, the dataset and analysis results of our study have been made publicly available online~\cite{replpack} in order to facilitate other researchers to replicate our study easily. We believe that these measures can partially alleviate this threat.

\section{Conclusions}
\label{sec:conclusions}
In this work, we conducted a mixed-methods study on code snippets in code reviews. We first analyzed the code review data mined from four most active projects of the OpenStack (Nova and Neutron) and Qt (Qt Base and Qt Creator) communities. We then conducted an industrial survey to get the practitioners' perspective on code snippets in code reviews. More specifically, we manually analyzed the extent of using code snippets, the reviewers' purposes of providing code snippets, the developers’ acceptance of code snippet suggestions, and the reasons why developers do not accept code snippet suggestions in code reviews through the repository mining study, and we explored the scenarios in which reviewers provide code snippets, the developers’ attitudes towards code snippets, and the characteristics of code snippets developers expect reviewers to provide in code reviews through the industrial survey study. 

According to the study results, code snippets are not frequently used in code reviews, and most of the code snippets in review comments are provided by reviewers. Reviewers use code snippets in code reviews with the aim of \textit{Suggestion} and \textit{Citation}, in which \textit{Suggestion} is the main purpose, and most developer would accept reviewers' code snippet suggestions. \textit{Difference in the opinions between developers and reviewers} and \textit{Reviewer's suggestion is flawed} are the main reasons why developers do not accept reviewers' code snippet suggestions. Reviewers tend to provide code snippets in code reviews \textit{when code is more illustrate than words}. Most developers hold positive attitudes towards provided code snippets in code reviews, and they expect that the provided code snippets can be \textit{understandable} and \textit{fitting into existing code}.

Based on the study results, we found that code snippets can be used as a special way to meet developers' information needs in code reviews, and appropriately using code snippets in code reviews will make the communication between developers and reviewers more effective. We suggested novel insights and future directions related to code snippets in code reviews for researchers. Meanwhile, we provided useful knowledge and information about using code snippets in code reviews, which can guide practitioners toward a more effective code review process.

In the next step, we plan to extend this work by studying code snippets in code reviews in a larger set of data sources, including pull request data from GitHub, more diverse projects from different communities (e.g., projects mainly written in Java from the Apache community). Besides, we intend to expand the scope of the industrial survey by including more participants from various development groups (e.g., inner source development). We also plan to explore the purposes of developers providing code snippets in code reviews through interviews and thus attain a more comprehensive understanding of code snippets in code reviews.

\section*{Acknowledgements}
This work has been supported by the National Natural Science Foundation of China (NSFC) under Grant No. 62172311 and the Special Fund of Hubei Luojia Laboratory. The authors would also like to thank all the participants of the online survey.

\section*{Data Availability Statements}
The data generated and analyzed during the current study is available in the Zenodo repository at \citep{replpack}.

\bibliographystyle{spbasic}
\bibliography{references}

\end{document}